\documentclass[a4paper,11pt,english]{article}
\pdfoutput=1
\usepackage{jinstpub} 
\usepackage{savesym}
\usepackage{amsmath}
\savesymbol{iint}
\savesymbol{iiint}
\usepackage{txfonts}
\restoresymbol{TXF}{iint}
\restoresymbol{TXF}{iiint}

\usepackage{verbatim}
\usepackage{hyperref}
\usepackage{pdfpages}
\usepackage{graphicx}
\usepackage{lineno}
\usepackage{epstopdf}
\graphicspath{{./fig/}}
\title{\boldmath Characterization of the Spontaneous Light Emission of the PMTs used in the Double Chooz Experiment}

\newcommand{\LN}{{\textit{LN~}}}
\newcommand {\an}{$\bar{\nu}_{e}$}

\affiliation[a]{Department of Physics, Tokyo Institute of Technology, Tokyo, 152-8551, Japan }
\affiliation[b]{Centro Brasileiro de Pesquisas F\'{i}sicas, Rio de Janeiro, RJ, 22290-180, Brazil}
\affiliation[c]{Max-Planck-Institut f\"{u}r Kernphysik, 69117 Heidelberg, Germany}
\affiliation[d]{III. Physikalisches Institut, RWTH Aachen University, 52056 Aachen, Germany}
\affiliation[e]{Physik Department, Technische Universit\"{a}t M\"{u}nchen, 85748 Garching, Germany}
\affiliation[f]{IPHC, Universit\'{e} de Strasbourg, CNRS/IN2P3, 67037 Strasbourg, France}
\affiliation[g]{University of California, Davis, California 95616, USA}
\affiliation[h]{Research Center for Neutrino Science, Tohoku University, Sendai 980-8578, Japan}
\affiliation[i]{Institute of Nuclear Research of the Russian Academy of Sciences, Moscow 117312, Russia}
\affiliation[j]{The Enrico Fermi Institute, The University of Chicago, Chicago, Illinois 60637, USA}
\affiliation[k]{Department of Physics and Astronomy, University of Alabama, Tuscaloosa, Alabama 35487, USA}
\affiliation[l]{AstroParticule et Cosmologie, Universit\'{e} Paris Diderot, CNRS/IN2P3, CEA/IRFU, Observatoire de Paris, Sorbonne Paris Cit\'{e}, 75205 Paris Cedex 13, France}
\affiliation[m]{Centro de Investigaciones Energ\'{e}ticas, Medioambientales y Tecnol\'{o}gicas, CIEMAT, 28040, Madrid, Spain}
\affiliation[n]{Columbia University; New York, New York 10027, USA}
\affiliation[o]{Universidade Federal do ABC, UFABC, Santo Andr\'{e}, SP, 09210-580, Brazil}
\affiliation[p]{Massachusetts Institute of Technology, Cambridge, Massachusetts 02139, USA}
\affiliation[q]{SUBATECH, CNRS/IN2P3, Universit\'{e} de Nantes, Ecole des Mines de Nantes, 44307 Nantes, France}
\affiliation[r]{Department of Physics, Drexel University, Philadelphia, Pennsylvania 19104, USA}
\affiliation[s]{Kepler Center for Astro and Particle Physics, Universit\"{a}t T\"{u}bingen, 72076 T\"{u}bingen, Germany}
\affiliation[t]{Argonne National Laboratory, Argonne, Illinois 60439, USA}
\affiliation[u]{NRC Kurchatov Institute, 123182 Moscow, Russia}
\affiliation[v]{Commissariat \`{a} l'Energie Atomique et aux Energies Alternatives, Centre de Saclay, IRFU, 91191 Gif-sur-Yvette, France}
\affiliation[w]{Universidade Estadual de Campinas-UNICAMP, Campinas, SP, 13083-970, Brazil}
\affiliation[x]{Department of Physics, Kobe University, Kobe, 657-8501, Japan}
\affiliation[y]{Department of Physics, Kansas State University, Manhattan, Kansas 66506, USA}
\affiliation[z]{Center for Neutrino Physics, Virginia Tech, Blacksburg, Virginia 24061, USA}
\affiliation[aa]{Department of Physics and Astronomy, University of Tennessee, Knoxville, Tennessee 37996, USA}
\affiliation[ab]{Department of Physics, Illinois Institute of Technology, Chicago, Illinois 60616, USA}
\affiliation[ac]{Department of Physics, Kitasato University, Sagamihara, 252-0373, Japan}
\affiliation[ad]{University of Notre Dame, Notre Dame, Indiana 46556, USA}
\affiliation[ae]{Department of Physics, Tokyo Metropolitan University, Tokyo, 192-0397, Japan}
\affiliation[af]{Hiroshima Institute of Technology, Hiroshima, 731-5193, Japan}
\affiliation[ag]{Tohoku Gakuin University, Sendai, 981-3193, Japan}

\newcommand{\nowatpucr}{Now at PUC University, R. Marqu\^{e}s de S\~{a}o Vicente, 225 - G\'{a}vea, Rio de Janeiro - RJ, Brazil.}
\newcommand{\nowatcenbg}{Now at CENBG, Bordeaux University, CNRS/IN2P3, F-33175 Gradignan, France.}
\newcommand{\nowathawaii}{Now at Department of Physics \& Astronomy, University of Hawaii at Manoa, Honolulu, Hawaii 96822, USA.}
\newcommand{\nowatific}{Now at Instituto de F\'{i}sica Corpuscular, IFIC (CSIC/UV), 46980 Paterna, Spain.}
\newcommand{\nowatmaryland}{Now at Department of Physics, University of Maryland, College Park, Maryland 20742, USA.}
\newcommand{\nowatkobe}{Now at Department of Physics, Kobe University, Kobe, 658-8501, Japan.}
\newcommand{\nowatprisma}{Now at Institut f\"{u}r Physik and Excellence Cluster PRISMA, Johannes Gutenberg-Universit\"{a}t Mainz, 55128 Mainz, Germany.}
\newcommand{\nowatelinp}{Now at ELI-NP, ``Horia Hulubei'' National Institute of Physics and Nuclear Engineering, 077125
Bucharest-Magurele, Romania.}

\author[a]{Y.~Abe,}  
\author[b,1]{T.~Abrah\~{a}o\note{\nowatpucr},}   
\author[c] {H.~Almazan,} 
\author[d]{C.~Alt,}  
\author[e]{S.~Appel,}   
\author[f] {E.~Baussan,}  
\author[d]{I.~Bekman,}  
\author[g]{M.~Bergevin,} 
\author[h]{T.J.C.~Bezerra,}  
\author[i]{L.~Bezrukov,}   
% ...

\collaboration{Double Chooz Collaboration}
\date{\today}
\emailAdd{roberto.santorelli@ciemat.es}

\abstract{During the  commissioning of the first of the two detectors of the Double Chooz experiment, an  unexpected and dominant  background caused by the emission of light inside the  optical volume has been observed. A specific study of the ensemble of phenomena called {\textit{Light Noise}} has been carried out in-situ, and in an external laboratory, in order to characterize the signals and to identify the possible processes underlying the effect. Some mechanisms of  instrumental noise originating from the PMTs were identified and it has been found that the leading one arises from the light emission  localized on the photomultiplier base and  produced by the combined effect of heat and high voltage across the transparent epoxy resin covering  the electric components. The correlation of the rate and the amplitude of the signal with the temperature has been observed. For the first detector in operation the induced background has been mitigated using online and offline analysis selections based on timing and light pattern of the signals, while a modification of the photomultiplier assembly has been implemented for the second detector in order to blacken the PMT bases.}

                              
\begin{document}
\maketitle
\flushbottom


\section{Introduction} 

Evidence of the contamination of the physics data by an  instrumental background arising primarily from light signals produced inside the photomultipliers  assembly has been reported  by many experiments over the  past years  \cite{SK1,SNO,Kamland,DB,Reno}. In some cases the instrumental background has been described  as a fast ($\approx{10-100}$ ns) flash of light  or, in other cases, as a train  ( $\approx{1-10}~\mu$s) of  pulses similar to the glowing in gas. 

It is possible that the spurious light emission is  detected by some of the surrounding or opposite tubes in such a way that the event can satisfy the  trigger logic and be recorded on disk. In some cases it is necessary to turn off permanently the most active PMTs in order to reduce the contamination of the physics sample with spurious events or to limit an undesirable trigger dead time \cite{SK2}.
A set of  cuts, typically based on the event topology, has  to be used to reject  the instrumental background offline, and the evaluation of the cut efficiency and acceptance, as well as the monitoring over time of the background stability in rate and amplitude, are then  required over the entire data taking. Additionally a visual scanning of the events was required on some occasions in order to reject the remaining background  \cite{SK1} contaminating the physics sample. The possible $flashing$ or light $glowing$ of the optical units  is  considered a potential concern for the future giant neutrino detectors equipped with several thousands of PMTs \cite{LBNE,JUNO}.   A potential correlation of   the ensemble of phenomena called {\textit{Light Noise}}  ({\textit{LN}})  with  temperature and high voltage has been sometimes evidenced \cite{STR,KamlandT}, however  poor information can be found  in literature and, to the best of our knowledge, no systematic study of the effect  or explanation of the light emission mechanism  have been published. 

During the preliminary tests of the Double Chooz far detector (FD),  the first one operational of the two  foreseen by the project,  an unexpectedly high light signal rate  was measured before the detector was filled with liquid scintillator.  Photon emissions, not evidenced at the time of the characterization of the PMTs in laboratory and not caused by light leaks, were detected when the PMTs  were switched on, with a rate correlated to their high voltage values. 
 None of the tubes was able to justify the total trigger rate observed, but each PMT was an effective emitter  with a rate of  the order of  $\approx0.1 - 1$ s$^{-1}$, with some units more active than the others.  The pulses were able to trigger the DAQ and contaminate the physics data sample. The effect remained substantially unchanged after the filling of the detector.

Detailed investigations, also with  laboratory tests on photomultipliers of the same type as the ones installed in Double Chooz, have been carried out in order to clarify the features of the light pulse and its production mechanism. It has been observed that  the light emission  was likely localized on the photomultiplier base, and it was produced by the combined effect of heat and high voltage across the electric components.  Currently more than 75 \% of the total triggers in the far detector are identified as spurious events produced by this instrumental background. As predicted by the tests in laboratory an overall increase over time of  the \LN   rate has been measured during the physics runs.  Evidences of the  correlation of the rate with the seasonal variation of the detector temperature  have been additionally found after three years of data taking, thus a temperature control system has been operated in the far detector in order to reduce such dependence.   
At the same time the  bases of the near detector (ND) PMTs have been covered with a black sheet of a radiopure and chemically neutral material  (Lumirror) that substantially reduced the impact of that background. 

The  characteristic features of the effect, the details on the modifications implemented to the PMT assembly and the cuts used to reject  the  background are reported in the following. The results of the  investigations can be particularly relevant  for other experiments that are known to use similar optical units   or base assembly \cite{Reno,Kamland,DBDP}.

\section{The Double Chooz  light detection system}

\begin{figure}[t!]
\begin{center}
  \includegraphics[height=.45\textheight]{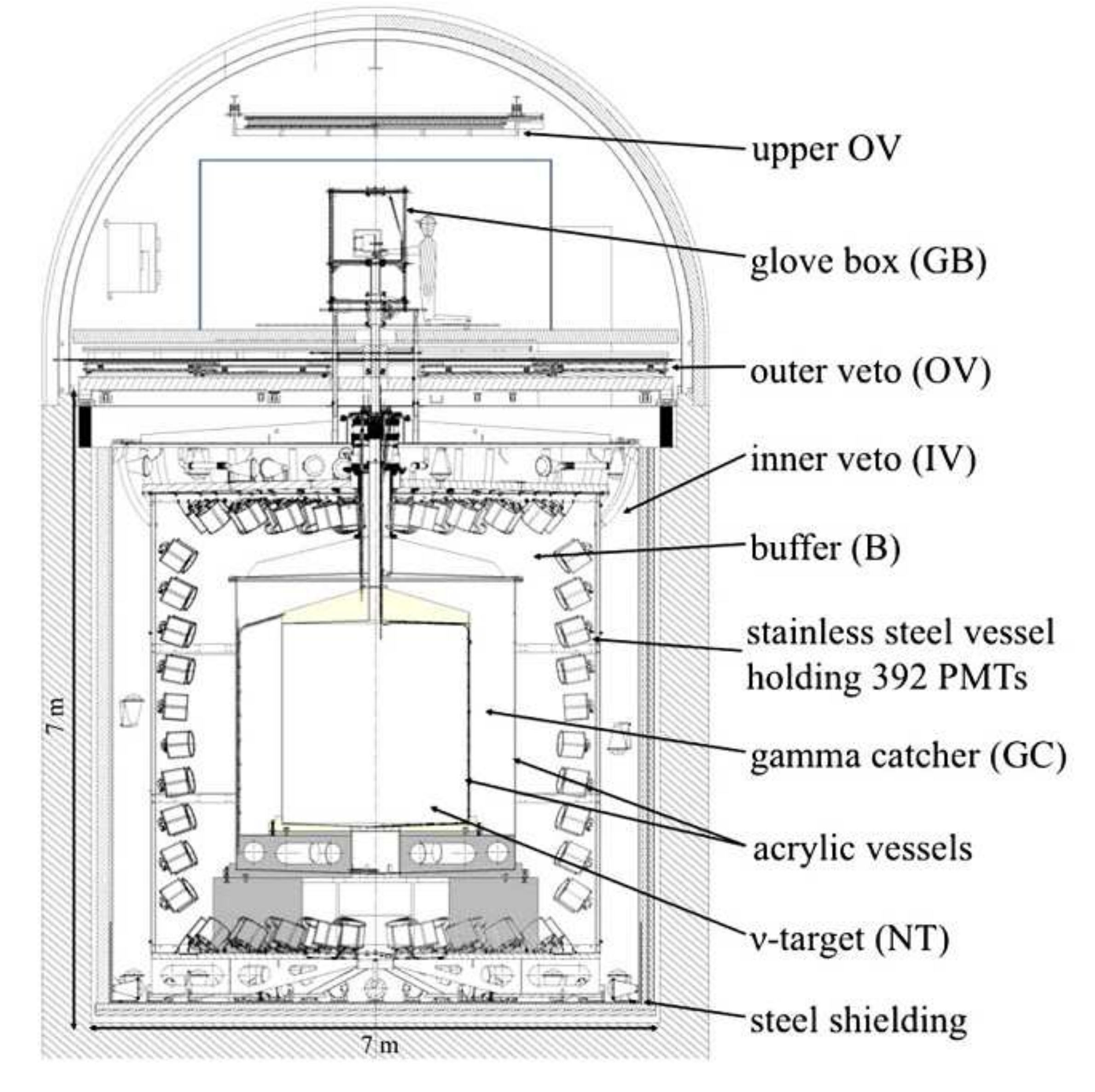}
  \caption{Cross section of the far detector.}
  \label{fig:detector}
  \end{center}
\end{figure}

Double Chooz is a reactor neutrino experiment which recently measured  the mixing angle $\theta_{13}$ through reactor antineutrino disappearance. The electron antineutrinos of energy $E_{\bar{\nu}_{e}}$,  produced by  the two reactor cores of the Chooz nuclear power plant in France, are detected through inverse beta decay (IBD): $\bar{\nu}_{e}$+$p\rightarrow e^{\textrm{+}}$+$n$ in hydrocarbon liquid scintillators that provide the free proton targets. The IBD signature is a coincidence of a prompt positron signal followed by a delayed neutron capture and the $\bar{\nu}_{e}$  energy   is reconstructed from  $E_{\textrm{prompt}}$, the positron visible energy ($E_{\bar{\nu}_{e}}\cong~E_{\textrm{prompt}} \textrm{+} 0.78$ MeV).  Results obtained with the first detector in operation,  located 1050 m from the two reactors,  show the reactor electron antineutrino disappearance to be consistent with neutrino oscillations. From a fit to the observed energy spectrum we found $\theta_{13}$ to be sin$^{2}2\theta_{13} $=$ 0.090^{\textrm{+}0.032}_{-0.029}$ \cite{DCG3}. The experiment continues to run and a near detector, placed at 400 m from the reactor cores and  almost identical to the far one, is now in operation and is expected to reduce the systematic uncertainties related to  reactor neutrinos (flux and energy)  and detection efficiency, achieving at least a precision  on sin$^{2}2\theta_{13}$ of $\sim$0.015  in three years of data taking \cite{DCG3}. 

\begin{figure}[t!]
\begin{center}
 \includegraphics[height=.38\textheight]{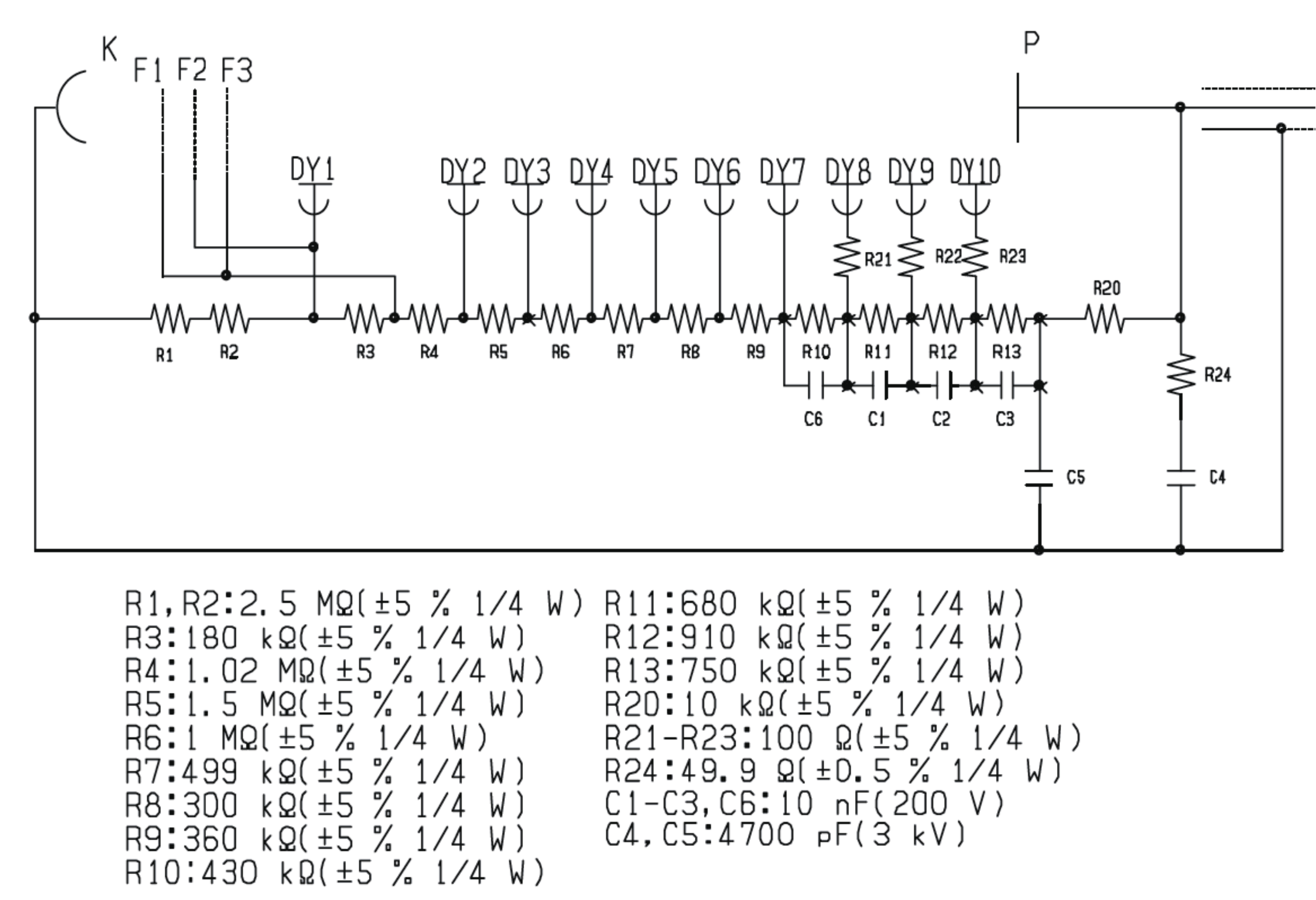}
 \caption{Schematic of the breeder circuit of the Double Chooz PMTs \cite{R7081}. }
 \label{fig:circuit}
   \end{center}
\end{figure}

The main bulk of the detector is made of four concentric cylindrical tanks with three central volumes, collectively called the inner detector (ID), optically  coupled. The  innermost of two acrylic vessels contains 10.3 m$^{3}$ of gadolinium-loaded (1 g/l) liquid scintillator (20 \% PXE, 80 \% Dodecane, PPO, Bis-MSB \cite{MPSci}) that acts as a  \an -target (NT, Fig. \ref{fig:detector}). The NT  is placed in the $\gamma$-catcher (GC), a 55 cm thick Gd-free liquid scintillator  (4 \% PXE, 46 \% Dodecane, 50 \% mineral oil, PPO, Bis-MSB) layer used to detect gamma rays escaping from the NT. The GC is surrounded by a 105 cm thick mineral oil layer (buffer) whose task  is to shield from radioactivity coming from  PMTs and surrounding rock. The light readout system is based on 390 10-inch PMTs that are installed on the inner wall of the stainless steel buffer tank and angled to maximize the light collection uniformity at the trigger stage \cite{DCDet}. Outside the ID, and optically separated from it, a  50 cm thick liquid scintillator  inner veto (IV) volume is equipped with 78 8-inch PMTs, and acts as an active shield to veto the interactions produced by cosmic muons or external radioactivity. The detector is surrounded by a 15 cm thick layer of demagnetized steel to suppress external gamma rays, and is covered by an outer veto system (OV) made of plastic scintillator strips modules to tag muons.

\begin{figure}[t!]
  \begin{minipage}[t!]{.25\linewidth}
 	   \includegraphics[height=0.24\textheight, angle=-90]{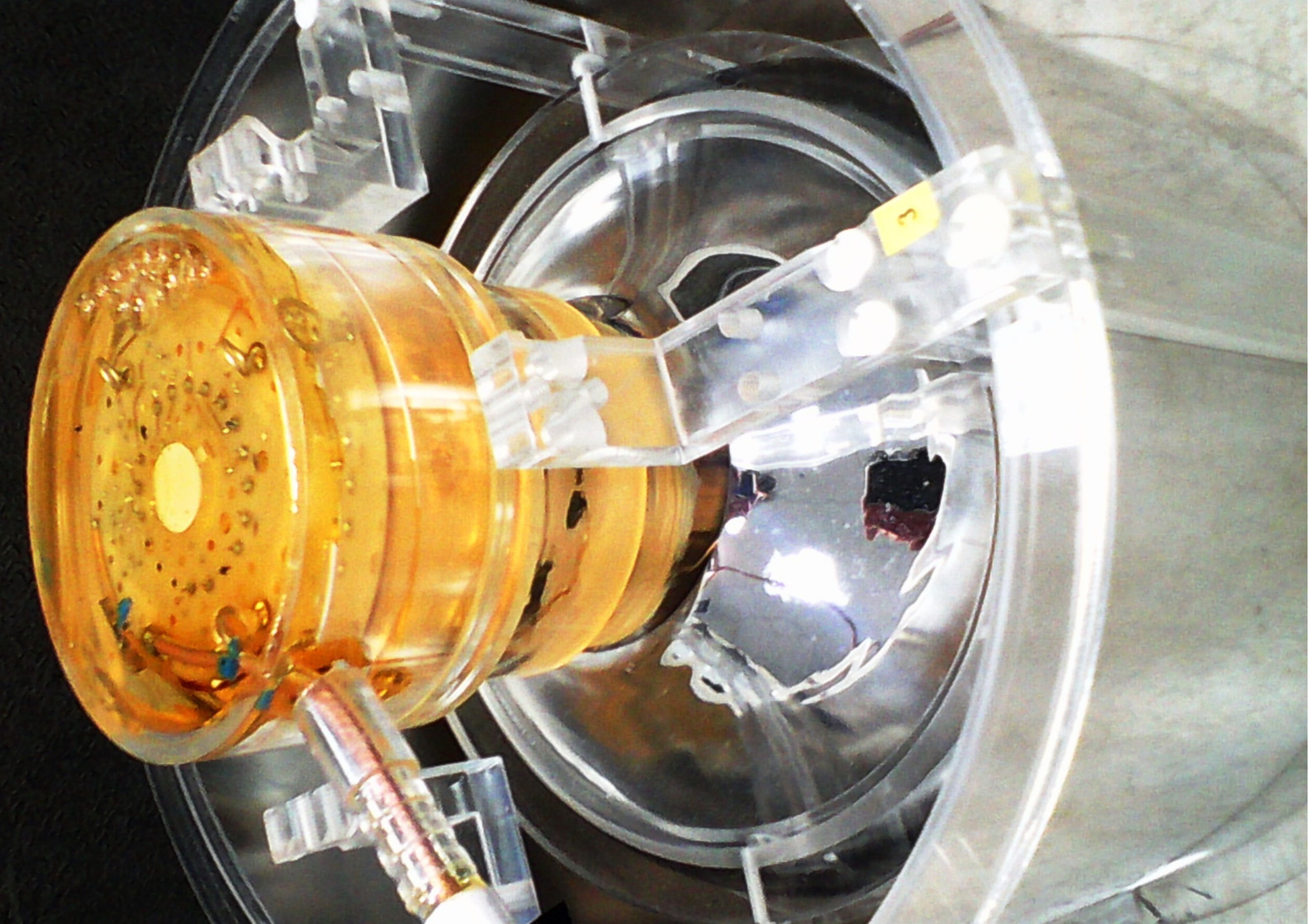}
  \end{minipage}
  \begin{minipage}[t!]{.75\linewidth}
  \begin{flushright}
     \includegraphics[width=0.9\linewidth]{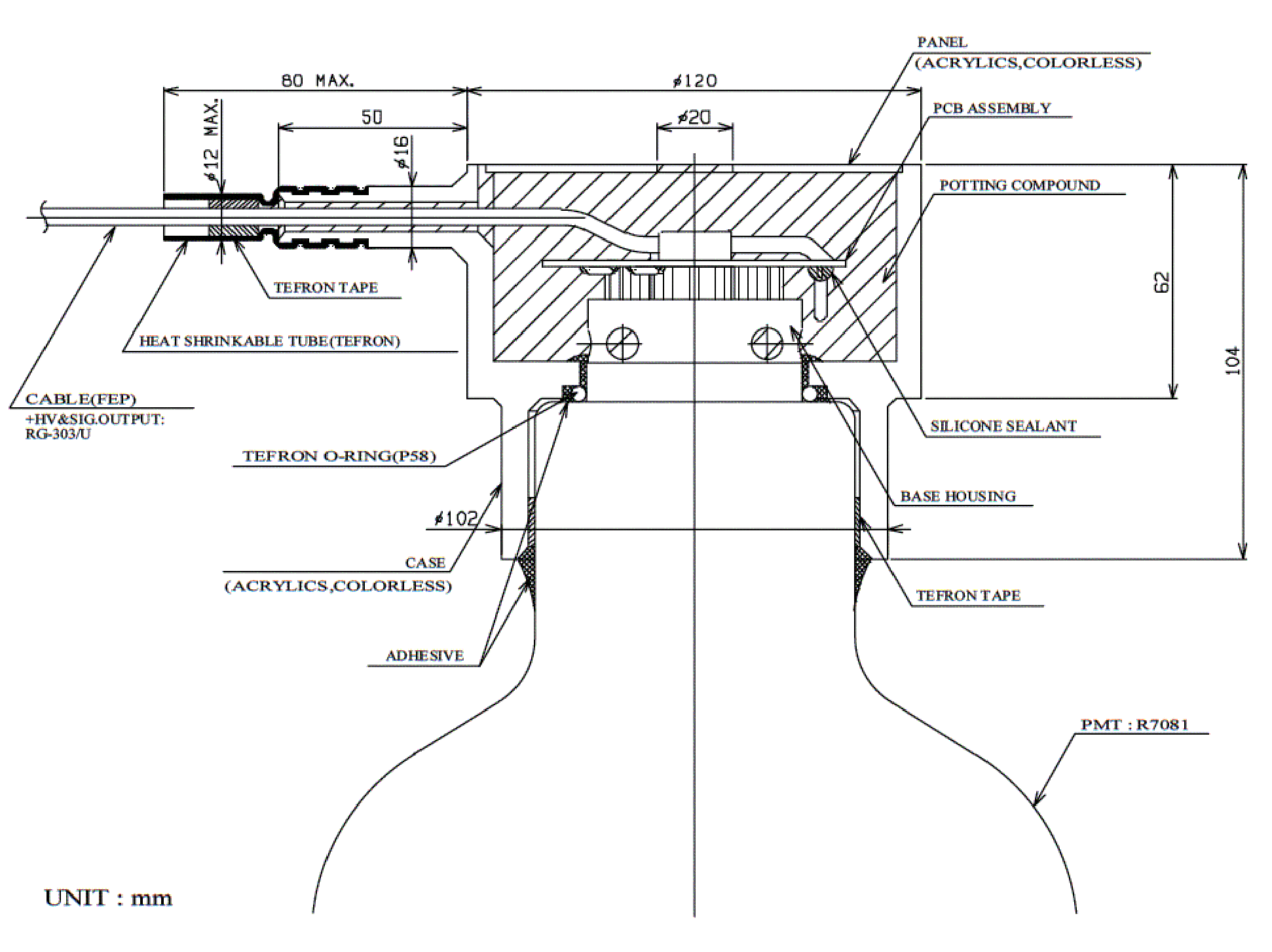}
     \end{flushright}
  \end{minipage}
  \caption{ Picture of the optical unit   (\textit{left}) and cross section of the base assembly  (\textit{right})  \cite{R7081} .}
 \label{fig:PMT_base}
\end{figure}

A set of 390 10-inch diameter low-background PMTs (R7081MOD-ASSY) produced by Hamamatsu Photonics \cite{R7081} is  used to detect the scintillation light produced in the ID. The 10 inch diameter low background PMTs are characterized by  good photoelectron separation  and timing resolution \cite{Matsubara,Bauer}. The PMT assembly is equipped with a breeder circuit (Fig. \ref{fig:circuit}) which accommodates a single cable readout\textsuperscript{\footnotemark[1]}. \footnotetext[1]{~Some ringing of the PMT signal was evident in case of very large light pulses \cite{DCDet}.} A round board and the cable end (Fig. \ref{fig:PMT_base}-{\it left}) are  soldered to the base electrodes.
The PMTs are operated with positive high voltage of 1320 V on average corresponding to a gain of $\approx 10^{7}$,  and a maximum  voltage difference of 520 V is generated between the cathode and the first dynode  of the PC board. About 0.05 W of heat dissipation on R1 and R2 (Fig. \ref{fig:circuit}) makes the PC board temperature locally slightly higher.
The entire PMT is submerged in an organic oil (buffer liquid) made of a mixture of n-alkanes and mineral oil  \cite{MPSci}. The maximum oil depth is 7 m and the maximum oil pressure on the PMT and its base is 0.6 atm.  The base circuit part is molded with  epoxy resin (main agent MG151+ hardner HY306 \cite{Pelnox}) to insulate the PCB and the electric parts from the oil as shown in Fig. \ref{fig:PMT_base}-{\it right}.  The  resin is contained in an acrylic case with a hole at the center of the end plate to allow for the volume change during the polymerization process of the resin. The choice of transparent acrylic and epoxy allows visual inspection of the sealing as well as of the electric components of the base circuit. The oil prevention treatment is similar to the  ones used for KamLAND, RENO and Daya Bay PMTs.

The whole PMT is surrounded by  a  cylindrical  (0.5 mm thick, 275 mm height and 300 mm inner diameter)  mu-metal magnetic shield. The dimensions of the shield and its relative position with respect to the PMT photocathode are optimized to maximize the shielding capability with the minimal loss of light collection \cite{Calvo}.  The distance between the top of the PMT and the shield edge is 5.5 cm. The inner wall of the mu-metal is a metal surface with a measured reflectivity of $\sim$ 60 \% at $\sim$~400 nm and, being almost completely specular, it acts as an effective light concentrator/collimator.

\section{Evidence of spontaneous emission of light by the Double Chooz PMTs}

\begin{figure}[t!]
\begin{center}
\includegraphics[height=4.5cm, width=7cm]{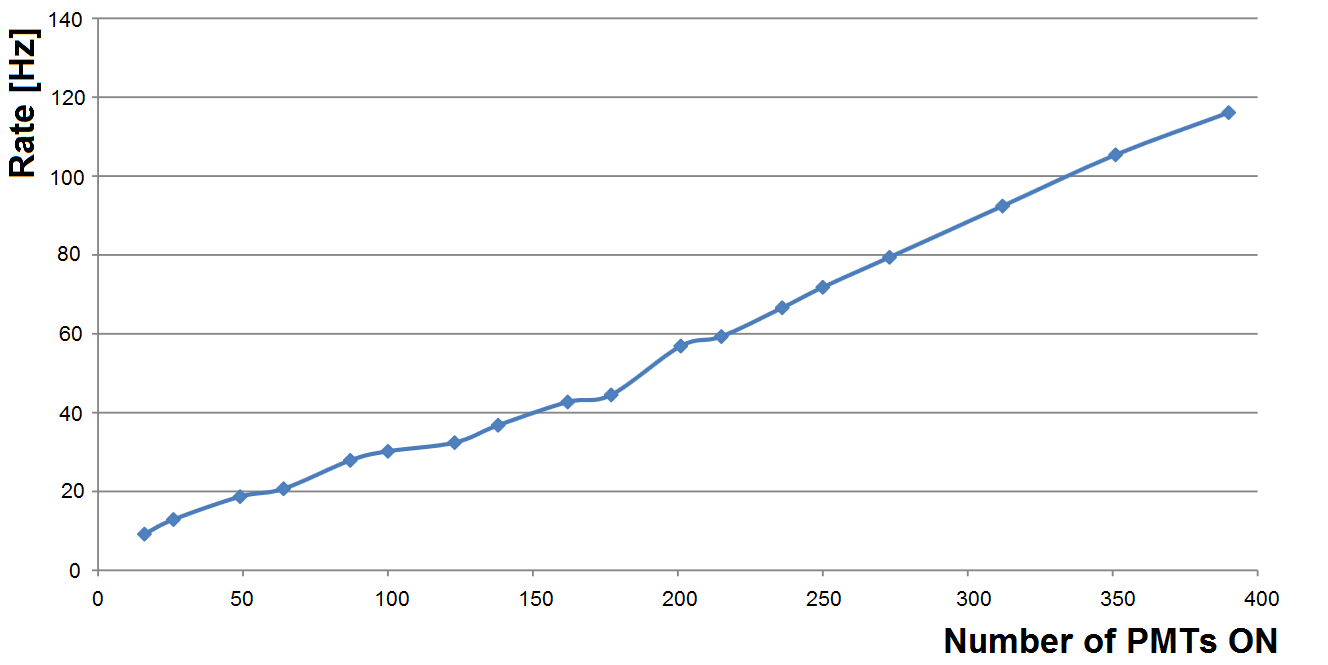}
\includegraphics[width=0.5\linewidth]{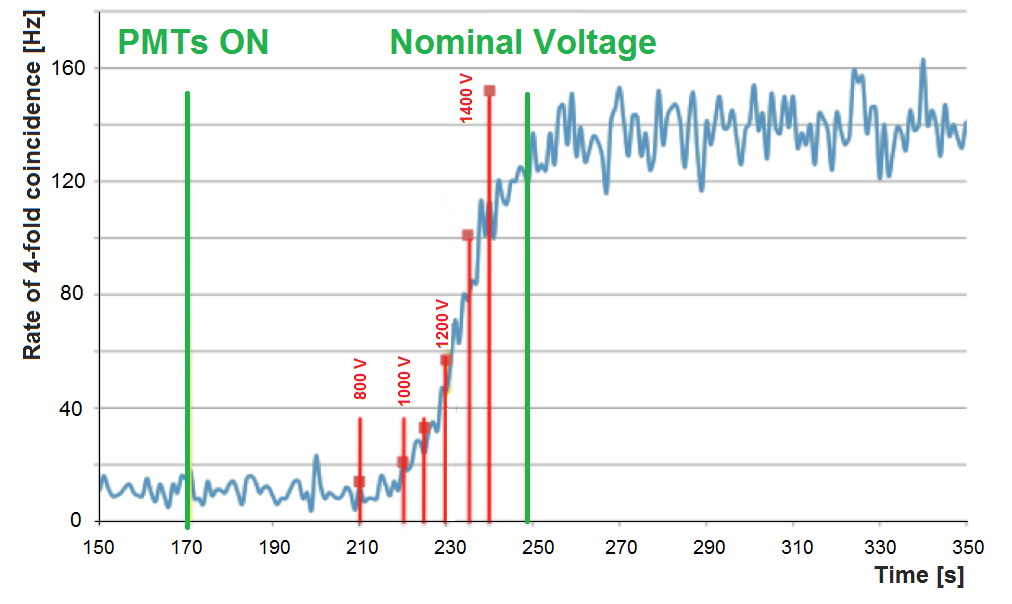} 
\caption{Coincidence rate measured with 16 monitor PMTs versus total number of PMTs on ({\it left}) and trigger rate progression as the high voltages of the monitored PMTs are ramped up ({\it right}) during the preliminary tests before the filling of the detector.  The rate is measured by the 4-fold coincidence. }
\label{fig:rate}
\end{center}
\end{figure}

During the  preliminary commissioning of the far detector and  before its filling with liquid scintillator, tests of the light detection system were carried out by taking data with a 16 channels portable readout system connected to 16 PMTs located on the bottom of the buffer vessel.  The PMT pulses were taken directly before the front-end electronics \cite{Sato} and the signal coincidence rate was recorded.  A  $\approx$~120 Hz    rate for a 4-fold multiplicity condition was measured with all 390 PMTs at 1000 V. The rate dropped to $\approx$~10 Hz powering  only the 16 monitoring PMTs, thus suggesting that the high coincidence  rate measured with all the PMTs switched on was produced by the emission of light inside the detector and was not caused by any light leaks\textsuperscript{\footnotemark[2]}.\footnotetext[2]{~Similar results were found for different sets of PMTs.} Additional scans were performed evidencing the correlation between the measured coincidence rate and the number of PMTs turned on (Fig. \ref{fig:rate}-{\it{left}}),  thus it was possible to exclude a possible malfunctioning of some specific units and explain the process as a general behavior of the PMT array.  \LN events appeared within few tens of seconds  after the voltage of the monitored PMTs was ramped up, and stayed stable over the following hours  (Fig. \ref{fig:rate}-{\it{right}}).
Subsequently an  even higher trigger rate was measured in operating conditions above the foreseen neutrino threshold (0.4 MeV) with the PMTs at the nominal voltages. 

\begin{figure} 
\begin{center}
\includegraphics[width=0.325\linewidth]{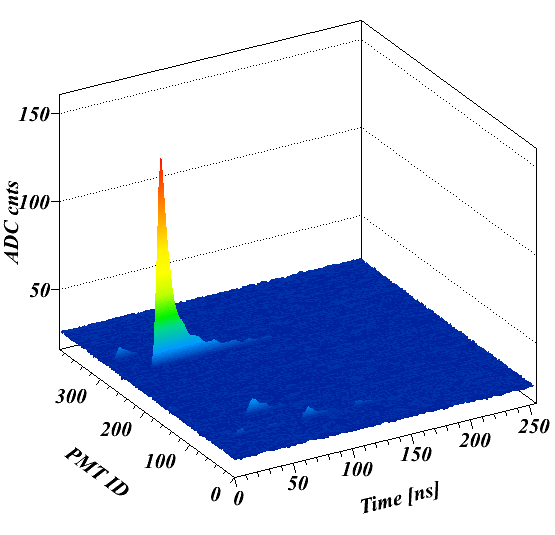} 
\includegraphics[width=0.325\linewidth]{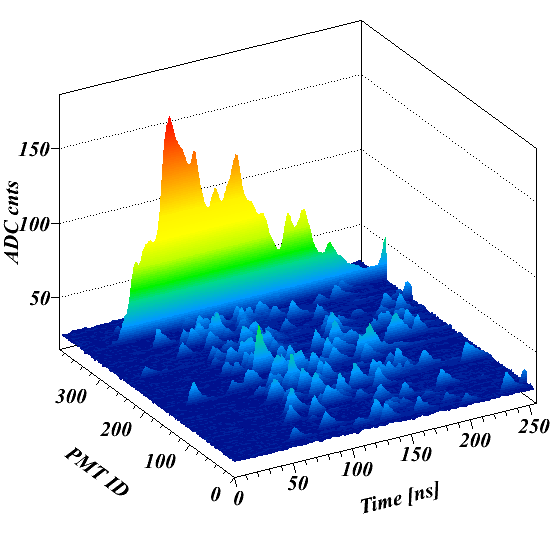} 
\includegraphics[width=0.325\linewidth]{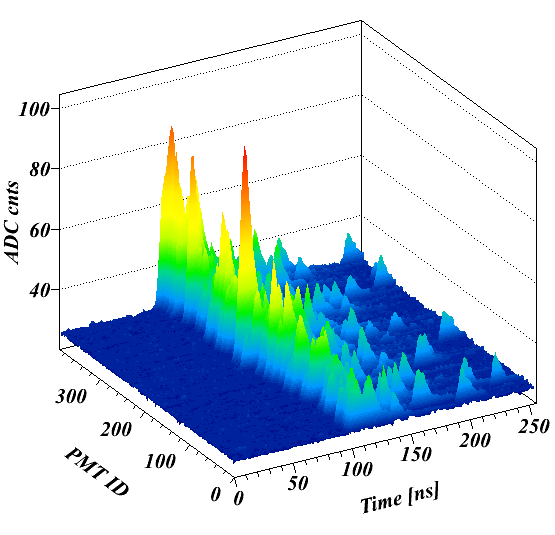} 
\caption{Time profile of the FD's PMT pulses detected for two typical \LN events (\textit{left} and \textit{middle}) compared with a  neutrino interaction (\textit{right}). Only in the last case the signals of the different PMTs are similar and  the time correlation between the arrival time of the photons is evident. }
\label{fig:LN}
\end{center}
\end{figure}

The effect remained substantially unchanged after the filling of the detector. A visual inspection of the waveform of those events excluded that the effect was produced by electronic noise, suggesting that the signal detected  appeared to be produced by light pulses, which spanned  in time from a few tens to hundreds of ns. In Fig. \ref{fig:LN}  two typical  events are compared with a neutrino interaction. In case of \LN,  larger pulses were detected by one specific PMT, on the contrary, in case of a neutrino interaction, similar light signals were detected by all PMTs with a time correlation featuring the time-of-flight of the scintillation photons.    
 Even though, on an event by event basis, the dominant charge  signal was  produced by one PMT, the light spread out inside the detector after several reflections,  triggering the DAQ and contaminating the physics data sample.  

\section{Laboratory tests and explanation of the effect} 
\label{SecTest}

\begin{figure}[t!]
\begin{center}
  \includegraphics[height=.3\textheight]{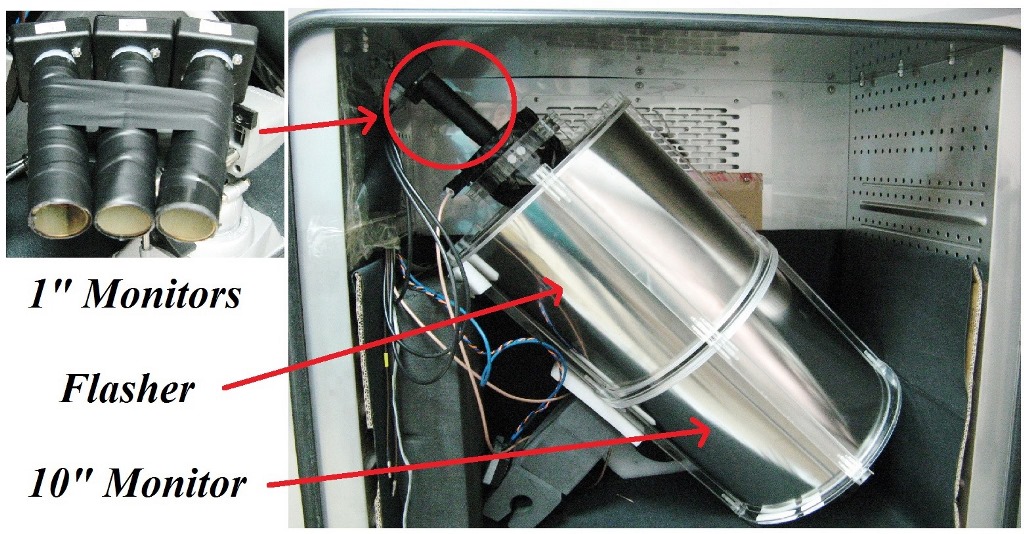}
  \caption{Experimental setup used for the tests in laboratory with the 4 monitor PMTs and the flasher  placed in a climate-controlled chamber.}
  \label{fig:fsetup}
  \end{center}
\end{figure}

The features of the light signal and the possible physical processes causing the emission of light was  studied in the laboratory with dedicated tests.  
 Data were acquired with three 1"  Hamamatsu R6095 PMTs placed on top of the  base (Fig. \ref{fig:fsetup}) of a 10" R7081 PMT\textsuperscript{\footnotemark[3]}.\footnotetext[3]{~The PMT has been previously identified as a flasher at room temperature.} The PMTs were located in a climate-controlled chamber (Binder MKF240) that was completely light-tight and allowed to carry out the test in a controlled environment of humidity and temperature. Another R7081 PMT\textsuperscript{\footnotemark[4]}\footnotetext[4]{~A negligible light emission rate has been measured for that PMT during preliminary tests.} was placed in front of the  tested unit in order to investigate the  light escaping its reflector shield.
Several runs have been taken in different configurations covering or uncovering  the base, the shield and the photocathode of the emitter  with a black sheet. While the main purpose of the measurement was the study of  rate, amplitude and pulse shape of the light  signal in different and controlled conditions, the small size of the 1" monitor PMTs placed directly on the base also makes the setup suitable for the identification of the light emission position.
 The glowing events are identified by the coincidence of the three 1" PMTs (200 ns  window) with a signal threshold corresponding to 1 photoelectron. An 8 $\mu$s window (1 $\mu$s pre-trigger) is then digitized  using a CAEN flash ADC (N6720) with 250 MHz sampling rate.  A LED placed inside the  chamber allowed to monitor the PMTs gain stability during the tests. Background runs for each configuration have been taken for reference by keeping the emitter switched off.

\subsection{Light emission features}

\begin{figure}
 \centering  \includegraphics[height=5.5cm, width=7.5 cm]{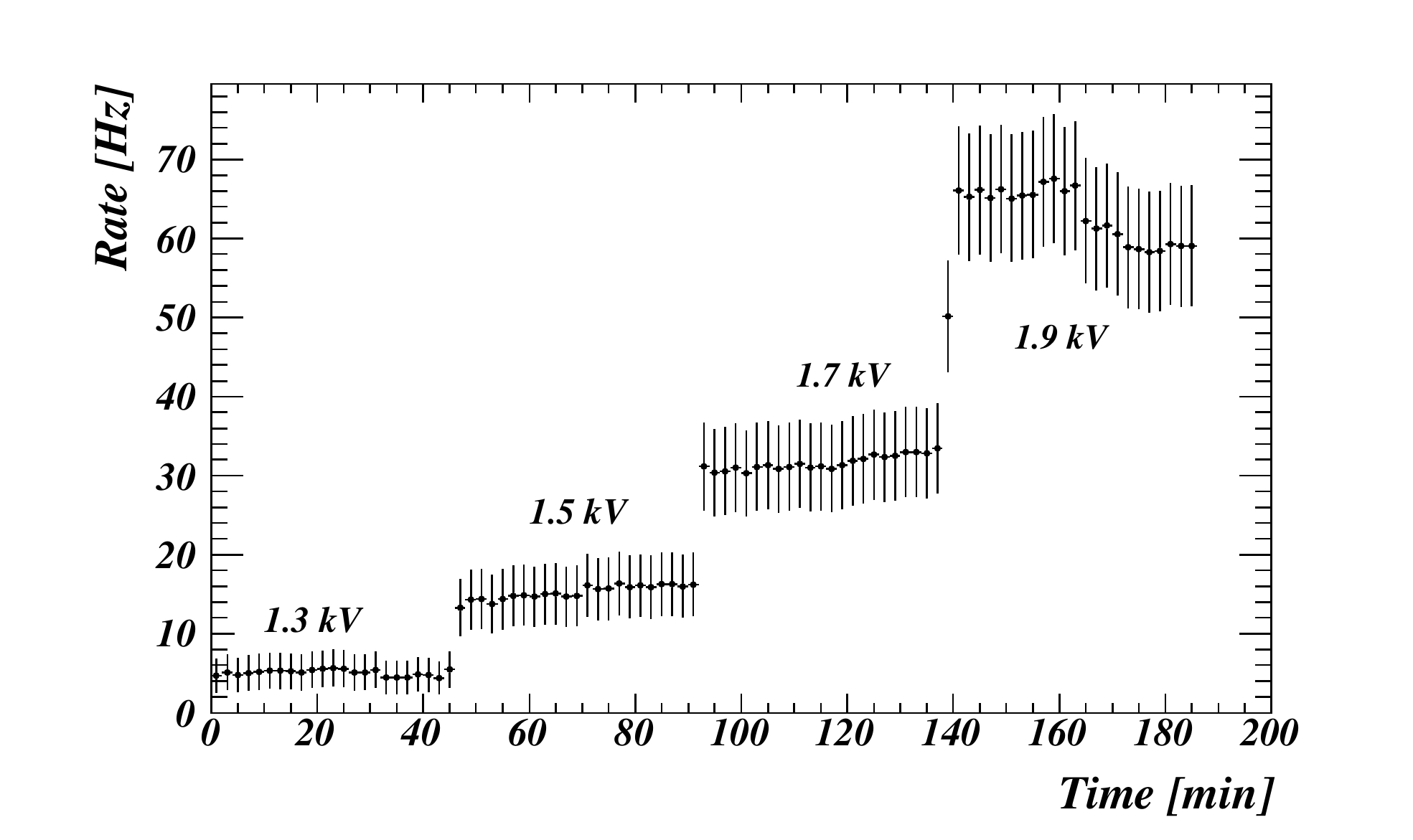}
 \centering  \includegraphics[height=5.3cm, width=7.5 cm]{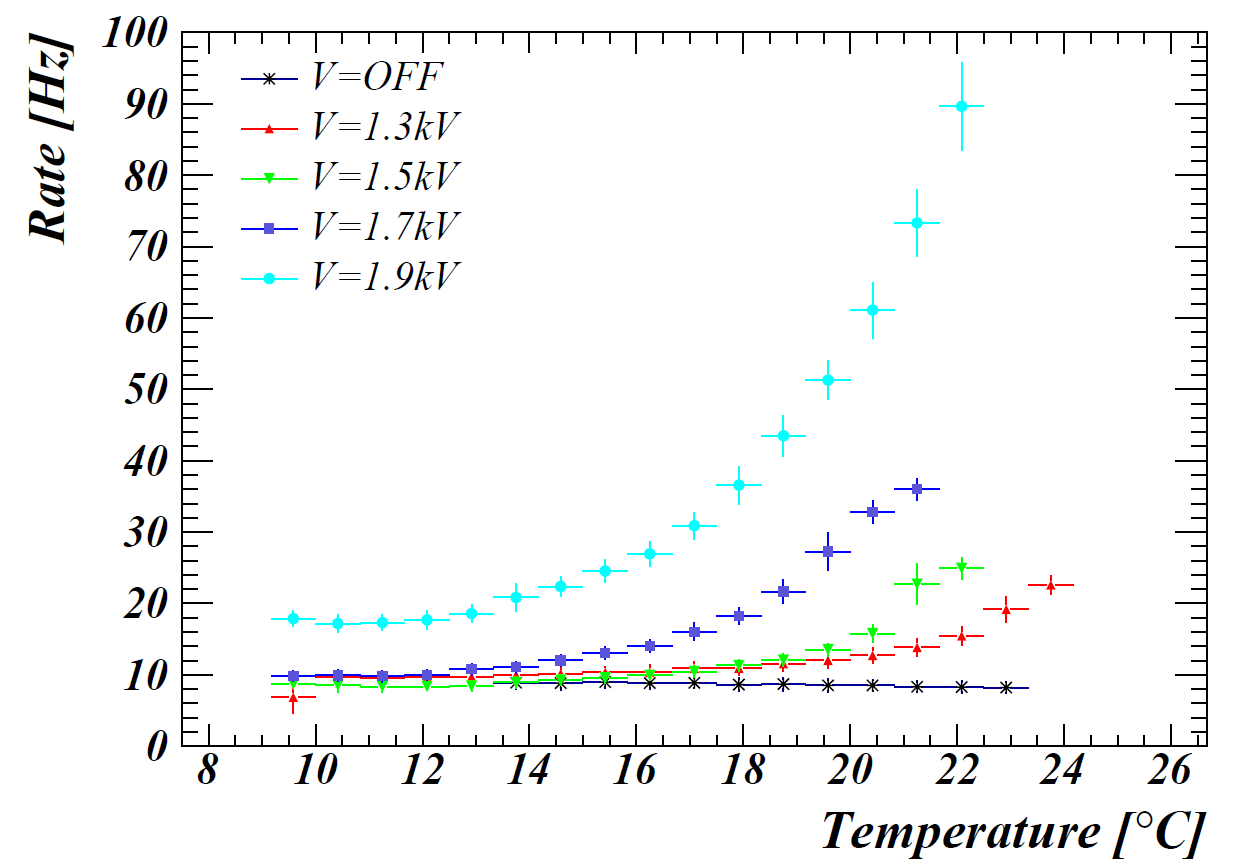}
  \caption{Variation of the light emission rate (3-fold coincidence) for different PMT high voltage values, keeping stable ({\it left}) or changing ({\it right}) the PMT temperature. A rise of the  temperature always corresponds to higher rates and bigger amplitudes of the detected signals.   }
  \label{fvar}
\end{figure}

Several measurements have been made
 changing the voltage supplied to the light emitter  and the temperature. In Fig. \ref{fvar} the prompt direct dependence of the measured coincidence rate with the increase of voltage and temperature is shown, confirming the observations already noted during the preliminary tests in the far detector. A  rise of the temperature  always produced an increase in the light production rate, however during the tests it was noticed that the corresponding  decrease of the temperature  was not able to lower the emission  in the same way. A rate hysteresis is thus produced moving back the temperature to the original value, with the frequency of the emission always higher after every  cycle.  In order to coherently  characterize the light emission, the  results presented in the following were obtained at constant values of temperature (T=25 $^{\circ}$C), voltage (V=1400 V) and humidity (60 \% relative value).

\begin{figure}[t!]
    \centering \includegraphics[height=5.5cm, width=7.5 cm]{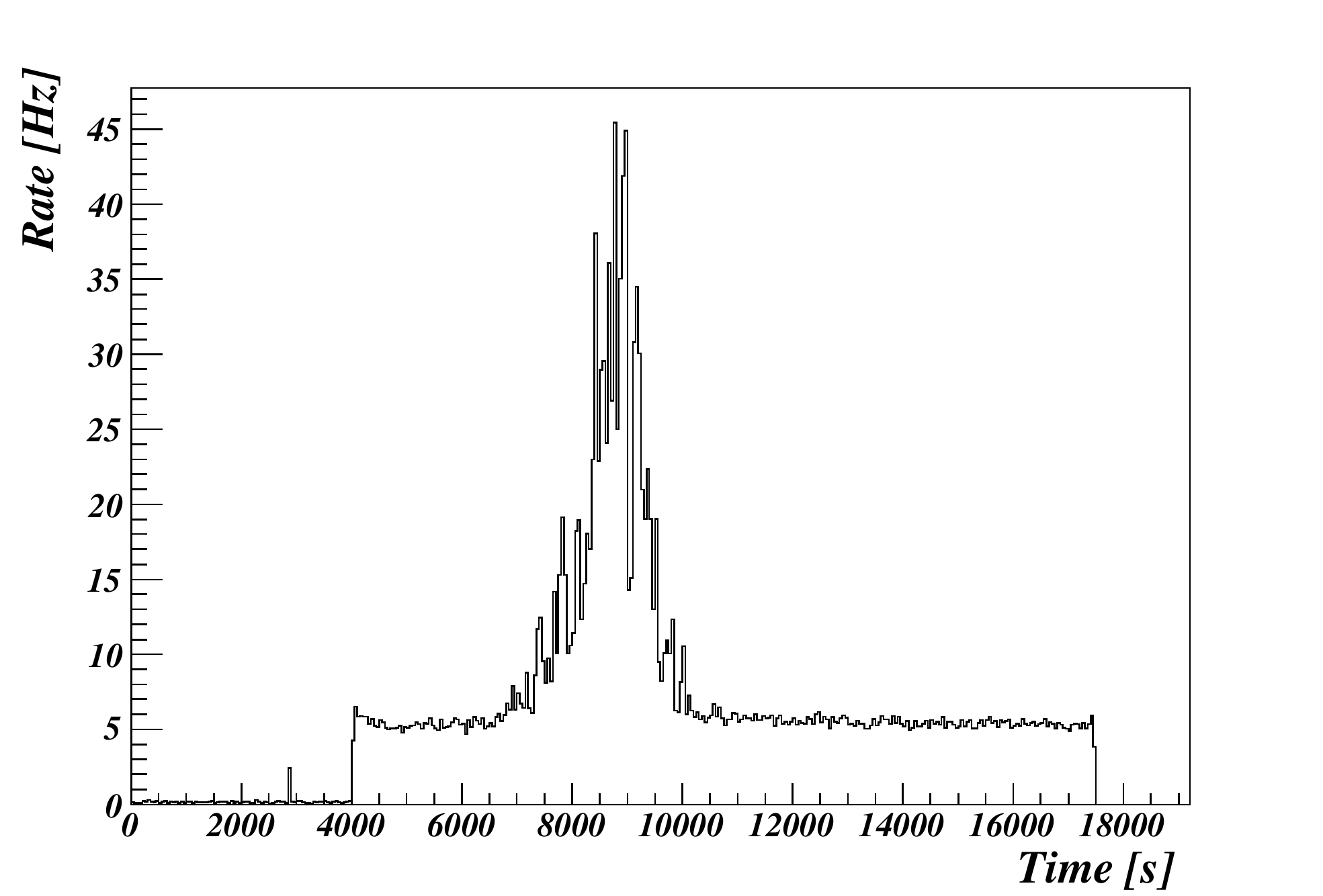}
    \centering \includegraphics[height=5.5cm, width=7.5 cm]{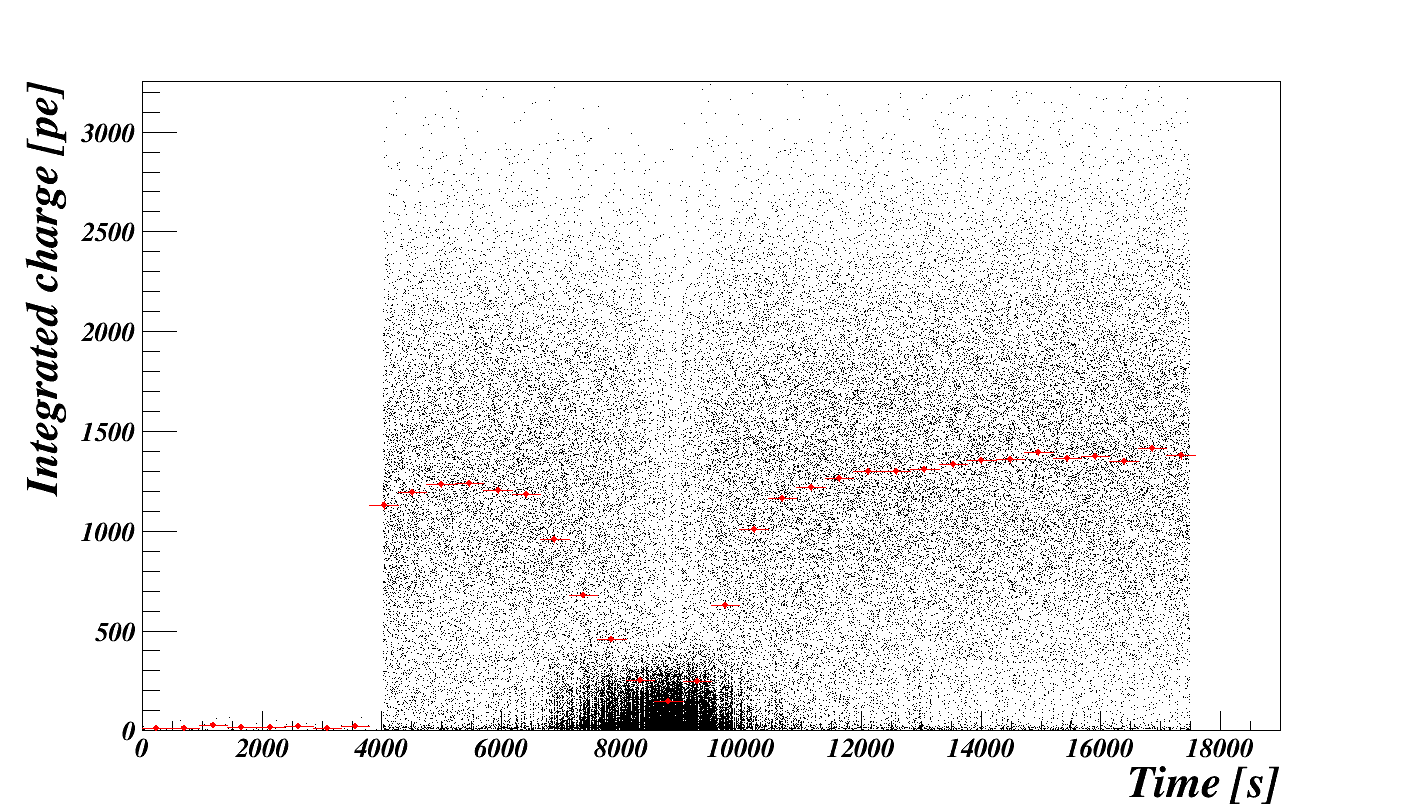}
  \caption{Variation of the monitoring PMTs coincidence rate during a \LN test in stable conditions ({\it left}). The coincidence  rate during the first 4000  s with the 10"  glower off are shown for reference. Evolution of the integrated charge  ({\it back dots - right}) and  variation of the overall mean value ({\it red dots - right}) as a function of the time. }
  \label{frate}
  \end{figure}
  
In Fig. \ref{frate}-{\textit{left}} the evolution of the trigger rate  is shown as a function of  time for a typical run taken during the tests in laboratory with the PMT in stable conditions. During the first part of the run (t \textless 4000 s) the glower was switched off in order to measure the background rate. Some triggers ($\approx$~0.1 Hz rate)  mainly given by cosmic rays and cross talk between the  channels have been recorded, however their impact  can be considered negligible for this analysis. Even though the PMTs have been kept in stable conditions, a significant variation of the rate is evident. An average trigger  rate of $\approx$ 6 Hz has been measured  shortly after the glower was switched on, however the rate rose to as much as 45 Hz for roughly a one hour transient and then went back down to $\approx$ 6 Hz again  without any apparent reason, thus evidencing a clear instability of the light emission process.

\begin{figure}
\begin{center}
  \includegraphics[height=.3\textheight]{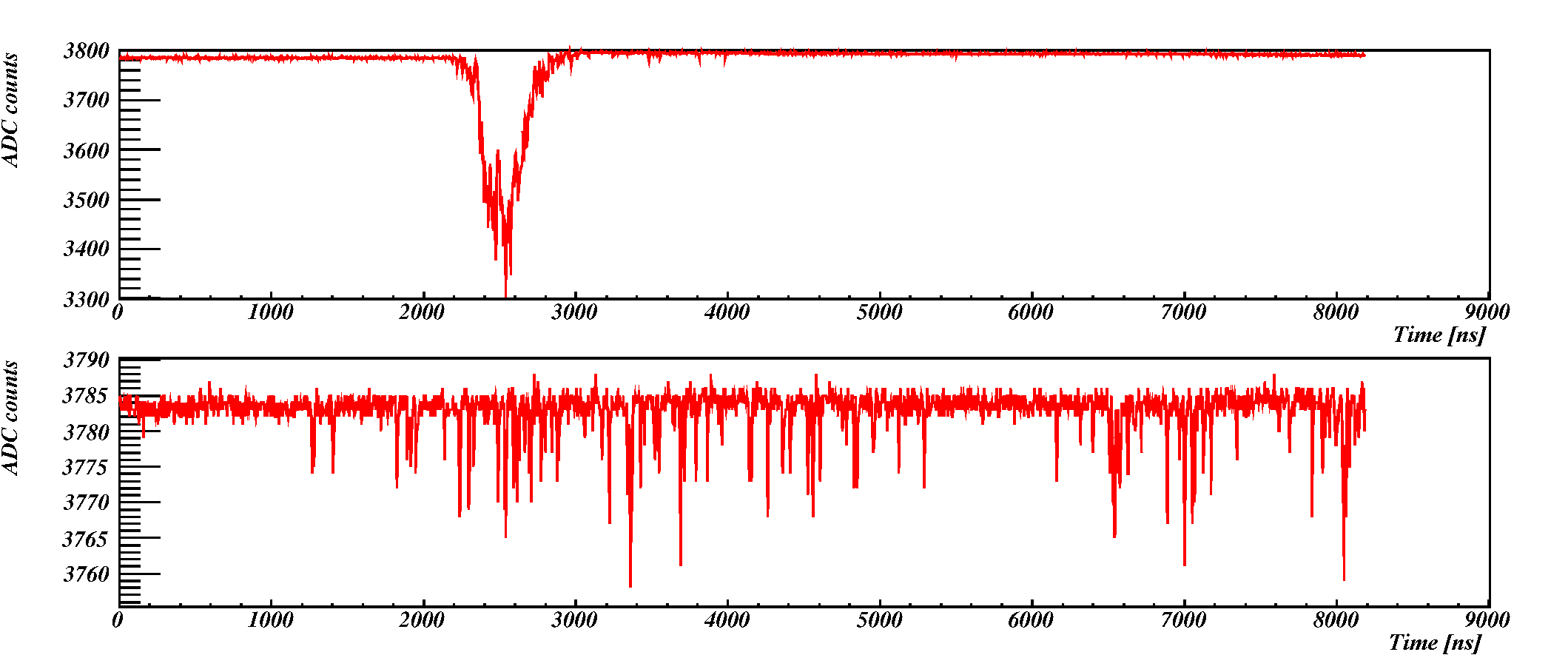}
  \caption{Different  \LN pulses detected during the tests in laboratory: while the first one is characterized by a bigger flash of light of some hundreds of ns, the second one is a train of small pulses of some $\mu$s. }
  \label{flabpulse}
  \end{center}
\end{figure}

\begin{figure}
\begin{center}
 \centering  \includegraphics[height=5.5cm, width=7.5 cm]{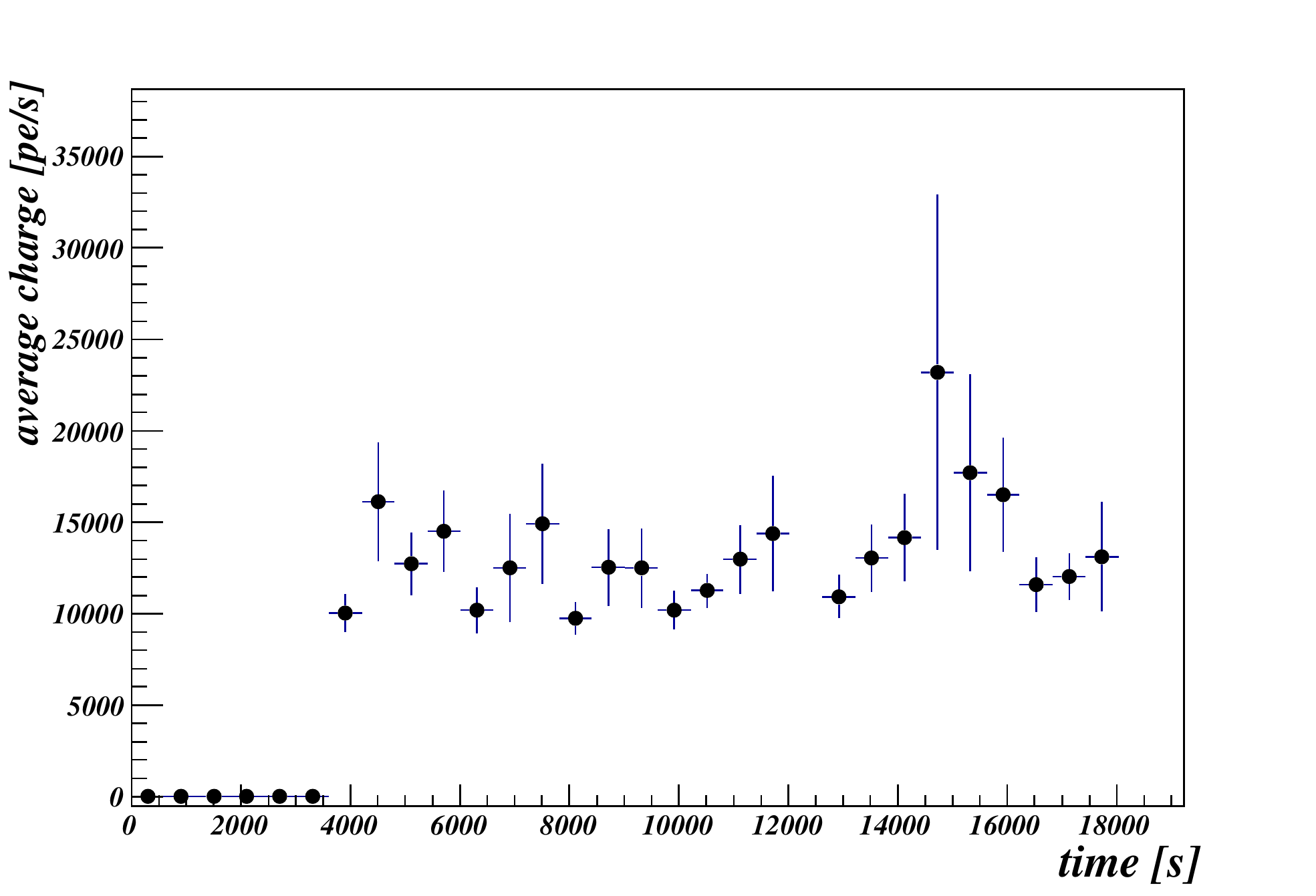}
 \centering  \includegraphics[height=5.5cm, width=7.5 cm]{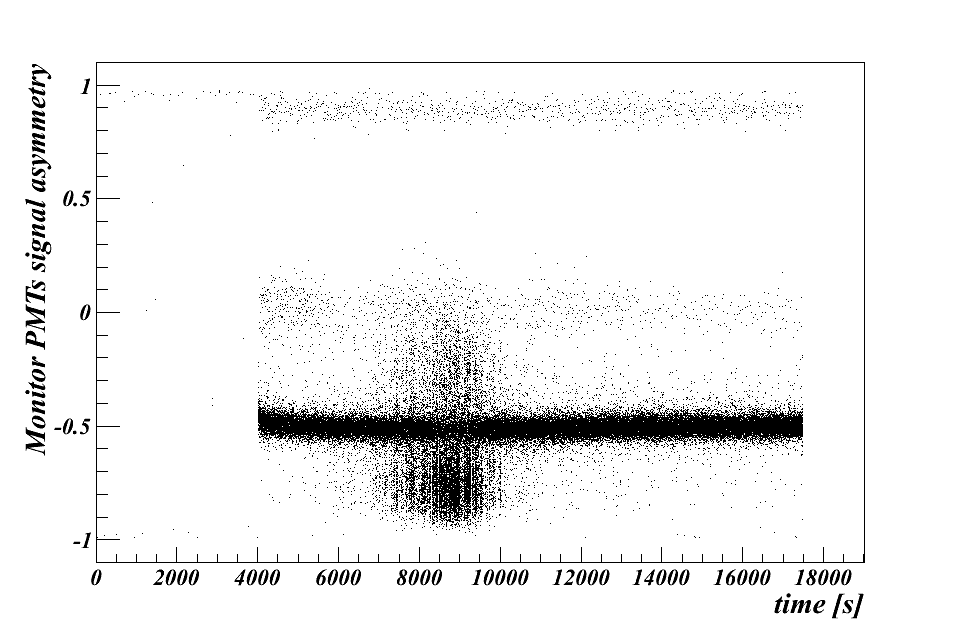}
  \caption{Average signal  detected in pe/s  ({\it left}) and evolution in time of the signal asymmetry (see text for details)
  between two monitor PMTs   ({\it right}) (same run of Fig. \ref{frate}). Multiple emission points can be identified on the base corresponding to the different asymmetry values. }
  \label{flabana}
  \end{center}
\end{figure}

The variation in time of the charge integrated in the same 200 ns digitization  window has been investigated in the regions evidenced by the different trigger rate values . In Fig. \ref{frate}-{\textit{right}} the  integrated signal in number of photoelectrons detected by one 1" monitor PMT  is plotted as a function of time. The increase in the rate is produced by signals with an amplitude smaller than the typical signal detected in the flat rate regions. In Fig. \ref{flabpulse} two typical waveforms are shown for events in the low  (first waveform) and high (second waveform) rate region. While the first one is characterized by the detection of hundreds of photoelectrons in 200-300 ns, the second event is given by the detection of a train of pulses of few ($\approx$ 1-2) photoelectrons over several $\mu$s. However, even though the  light pulses have widely different features, the integral signals obtained by summing up the total charge detected on large  ($\approx$ 1 s)  time scale is similar (Fig. \ref{flabana}-{\textit{left}}).  The average number of photolectrons detected per second seems to be constant and independent of the type of light pulses detected, thus suggesting a possible common  physical process that releases the same stored energy through  complementary  light production mechanisms. 
The analysis of the signal asymmetry  ($\textrm{S}_{\textrm{PMT1}}-\textrm{S}_{\textrm{PMT3}}$)/($\textrm{S}_{\textrm{PMT1}}\textrm{+}\textrm{S}_{\textrm{PMT3}}$) between the two  monitor 1" PMTs on the sides evidences different constant ratios  (Fig. \ref{flabana}-{\textit{right}}), suggesting the presence of different emission points on the base each of which characterized by a peculiar light signal in terms of shape, rate and amplitude.

Additional  tests carried out with other R7081 PMTs, initially not identified as flashers after tests at room temperature, showed that a light emission can be induced by raising  the temperature and/or the voltage to values higher than the normal operating conditions, thus suggesting that the light emission is a general effect that can characterize  any Double Chooz optical unit. At the same time  further tests carried out with a naked base not covered by epoxy did not evidence an emission of light in any condition. This excluded the possibility that the \LN was simply caused by thermal emission and suggested a  central role of the epoxy in the light production mechanism.

\subsection{Epoxy investigation}

\begin{figure}[t!]
  \centering \includegraphics[height=5cm, width=7.cm]{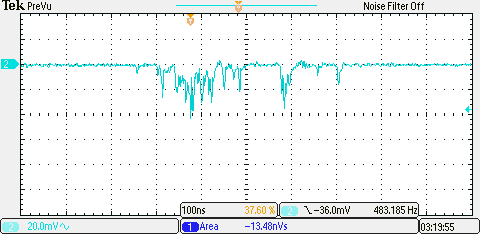}
  \centering \includegraphics[height=.22\textheight,angle=0]{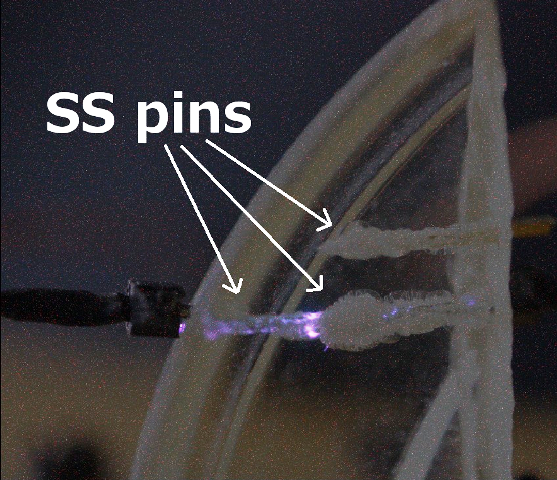}
  \caption{Typical signal detected with the epoxy sample in the electric field ({\it left}) and light emission recorded in dark conditions by the high sensitivity camera  after 5 minutes exposure  ({\it right}) with T$\approx$~50$^\circ$C (the epoxy sample image has been overlapped to the light recorded in dark conditions). Three metal pins at different potentials have been inserted in the sample, the light emission is evident only on the cathode pin (see text for details).  }
  \label{fsample}
\end{figure}

More detailed investigations on the behavior of the epoxy used to cover the PMTs base have been carried out in the laboratory. Two pins (0.6 mm diameter) have been connected to a power supply and inserted ($\approx$ 5 mm distance) in a small piece of epoxy ($\approx$ 10 g). The signal coincidence rate of two 1" PMTs, of the same type as used in Fig. \ref{fig:fsetup}, has been used to monitor the emitted light. A ceramic heater has been placed underneath the epoxy sample in order to raise its temperature. While no emission of light  has been observed without field between the pins, light signals have been detected with HV values of the order of 1 kV, thus confirming the central role of the epoxy in the production of light. Events similar to the ones detected during the tests of  the R7081 PMTs  have been detected (Fig. \ref{fsample}-$left$), with shapes  related to the distance between the pins and pin holes sizes. 
Additionally a clear dependence of rate and amplitude of the signals  on HV  and temperature has been evidenced as in the previous tests. A major rise in the temperature of the sample ($\approx$ 50$^\circ$C average value measured with a thermographic camera) and  HV ($\approx$ 4-5 kV) of the sample  made possible to take pictures of the light emission  with a high sensitivity digital camera on several minutes exposure. In Fig. \ref{fsample}-$right$ the light pattern recorded by the camera (almost blue or violet) has been superimposed on the image of the epoxy sample and the HV pins. A clear emission point can be noticed at the edge of one pin, while the rest of the pattern is mainly dominated by the light diffracted on the cracks in the epoxy holes. The emission of light is localized close to the electrode with the lower potential while it seems to be absent (or too low to be detected) on the other one. That is also the case switching the polarity between the opposite pins: after a short transient (few minutes) when the light is visible on both pins the luminosity near the anode decreases and eventually disappears. Also, more intense light emissions have been observed when the epoxy and the pins are closer (or in direct contact).

\begin{figure}[t!]
\begin{center}
  \includegraphics[height=5cm, width=7.cm]{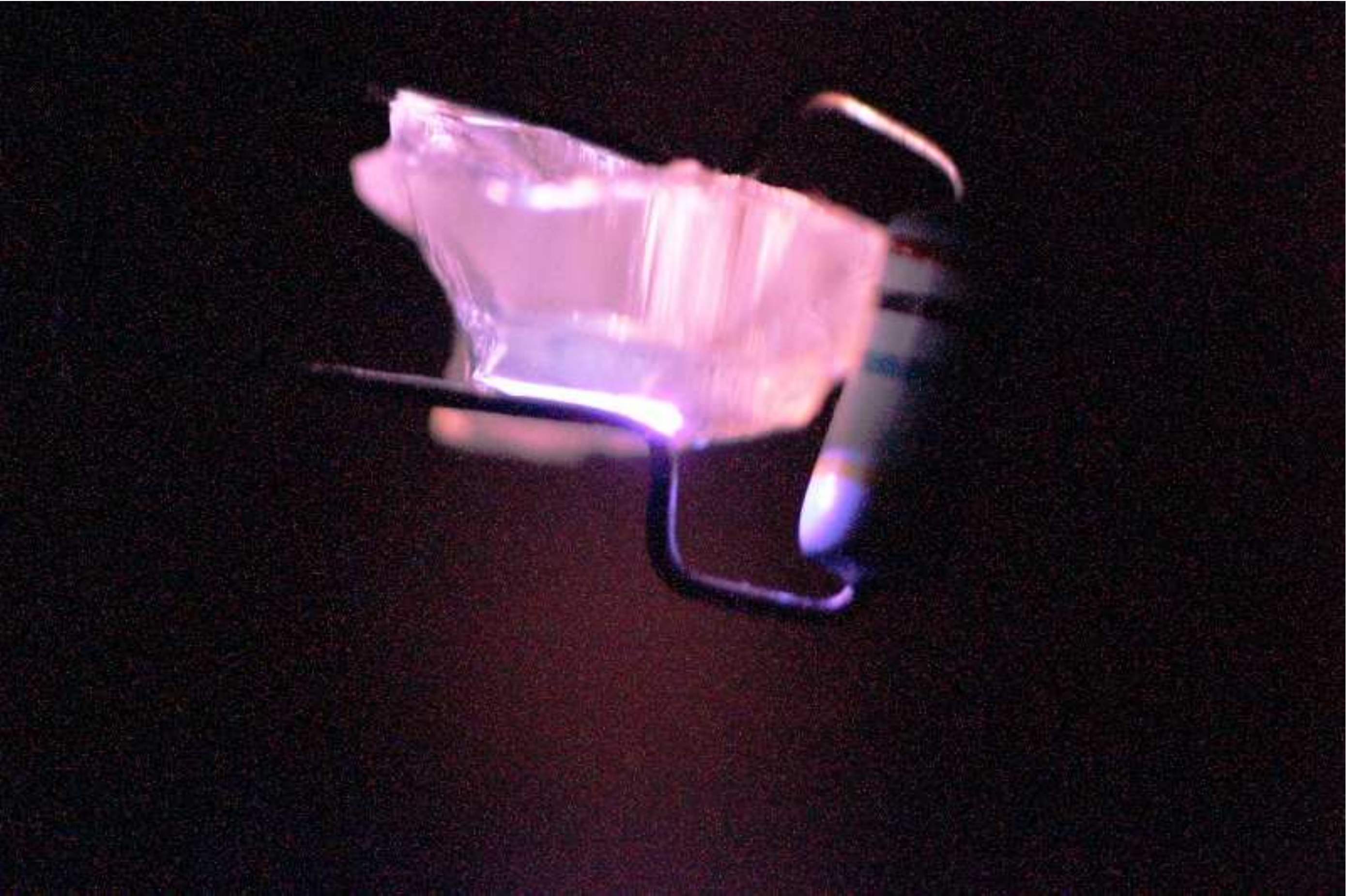}
   \includegraphics[height=5cm, width=7.cm]{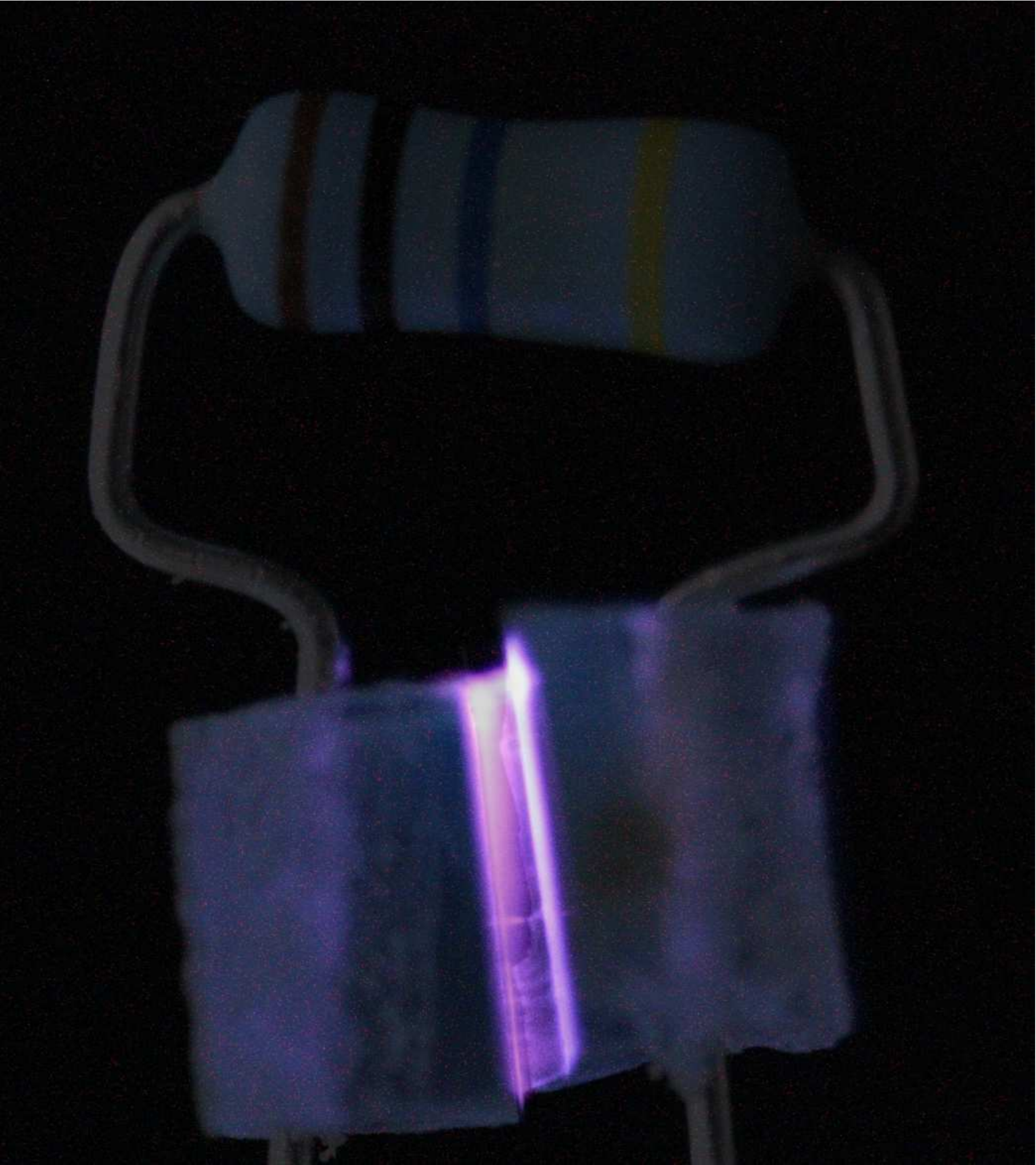}
  \caption{Glowing of epoxy samples placed on a resistor recorded with the camera in dark conditions.}
  \label{frsg}
  \end{center}
\end{figure}

In order to fully prove that the light emission  of the PMTs can be produced by the combined effect of heat and high voltage on the epoxy glue covering the base circuit,
a small sample has been placed between the two pins of a 10 M$\Omega$ resistor similar to the ones used on the Double Chooz PMTs base, that also acts as heater for the epoxy when a current of some mA flows in the resistor. In Fig. \ref{frsg}-{\it left} the image recorded by the camera in dark conditions is shown. A clear bright spot is evidenced where the cathode pin is in contact with the epoxy, while the illumination of the sample body is probably given by the  diffracted light. No light emission  has been recorded  by the camera in dark conditions after removing the epoxy or isolating the metal pins with black tape, thus preventing the direct contact of the epoxy with the electrodes.  
A similar glowing emission has been observed with two different pieces of epoxy in contact with the resistor pins and separated in the middle by a small gap (Fig. \ref{frsg}-{\it right}). Light production, probably given by corona discharge in air caused by the polarization of the two samples, is evident while no electric discharge is observed putting the pins at the possible minimum distance.  On the contrary  the light production point moves back to the cathode pin location when putting the two epoxy pieces in direct contact.
Similar results have been obtained carrying out the same tests with other different types of transparent epoxy commercially available (Epo-Tek\textsuperscript{\textregistered} 301-2 \cite{Epotek}, Araldite\textsuperscript{\textregistered} 2011 \cite{Arald}).

The previous  results fully proves that the epoxy  used to cover  the electronic components can cause the emission of light under certain conditions of voltage and temperature, thus explaining the light emission from the Double Chooz PMTs base.  A possible simple explanation of the  effect can be given considering that, due to the 
polarization of the epoxy, a strong field is present around the electrodes. According to that scenario
photons can be produced by the glowing of the air trapped between the epoxy and the pins through a corona effect. It is possible that the vapor pressure of the epoxy components could also play a role causing the corona. The possible glowing of the gas trapped at the interface between insulating layers with different dielectric properties is a well known phenomenon in  industry, and  procedures to remove air from  epoxy or urethane mixtures are typically suggested before the encapsulation of electronic components \cite{crossl}.

\section{Characterization  of the Light Noise during the commissioning of the far detector} 
\label{Sec:Comm}

\begin{figure}[t!]
\begin{center}
\includegraphics[width=0.82\linewidth]{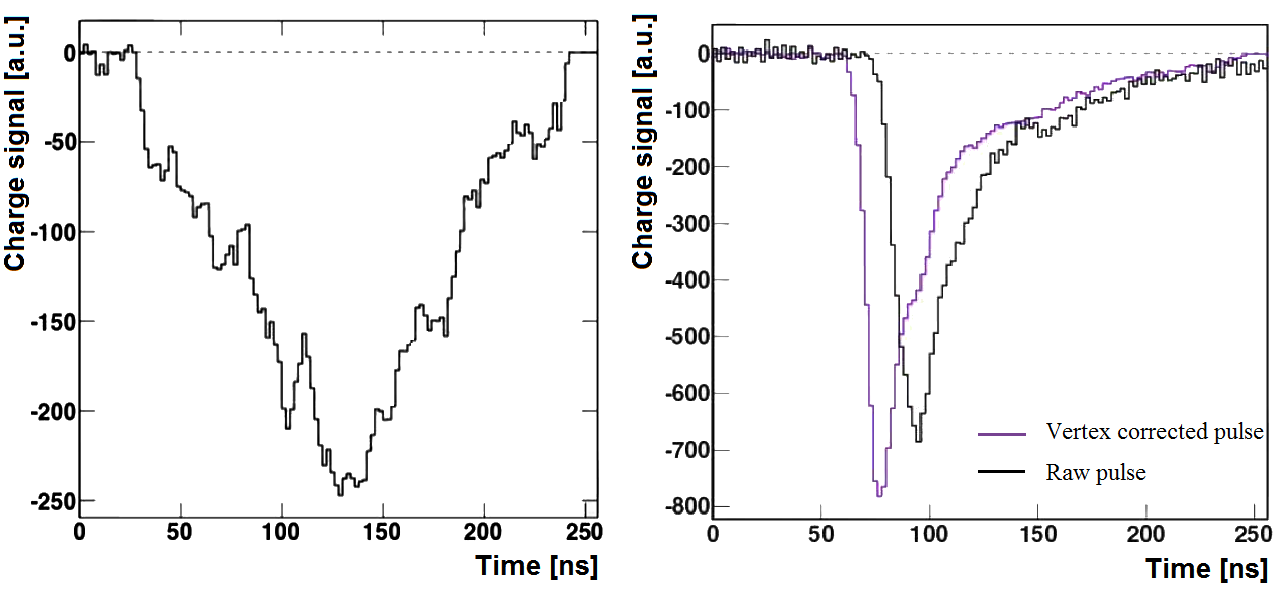}
\caption{Summed waveforms  recorded with the flash ADCs for  a \LN  event ({\it left})  compared to a typical signal produced by the liquid scintillation ({\it right}). The shapes for the purple line have been corrected taking into account the estimated vertex position and the photons time of flight.}
\label{fig:pulses}
\end{center}
\end{figure}

The Double Chooz Flash ADCs record the complete waveforms in a 256 ns digitization  window with 2 ns sampling, thus allowing the full comparison between spurious and scintillation emissions  in terms of pulse shape. The  \LN signals are typically characterized by a  quite irregular shape with a relatively slow rise time ($\approx$ 70 ns, Fig. \ref{fig:pulses}-{\it{left}}) compared to the well known  smooth shape with short rise time ($\approx$ 10 ns, Fig. \ref{fig:pulses}-{\it{right}}) typical of the scintillator.

As evidenced in the previous section  one of the possible kinds of \LN emission is in form of a long train of small pulses at the level of the single photoelectron. Thus, in this case, it is possible that the baseline measured in the pre-trigger region (first 50 ns of the digitized signal)  significantly differs from  the usual value since small pulses, not producing any trigger, can be present. Additionally a single spurious light emission can  produce one or even two triggers resulting in correlated background. 

\begin{figure}
\begin{center}
\includegraphics[width=0.5\linewidth]{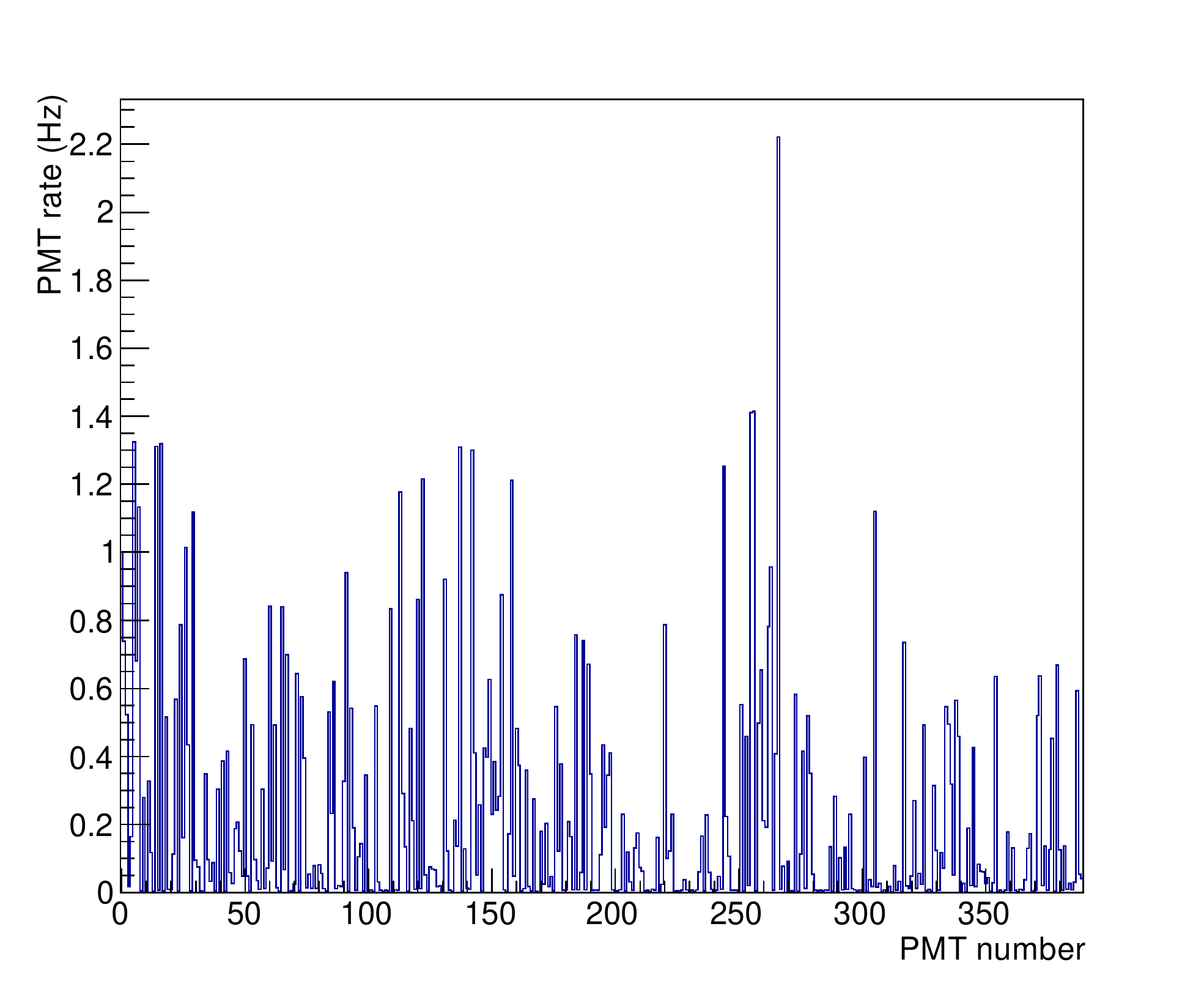}
\includegraphics[width=0.49\linewidth]{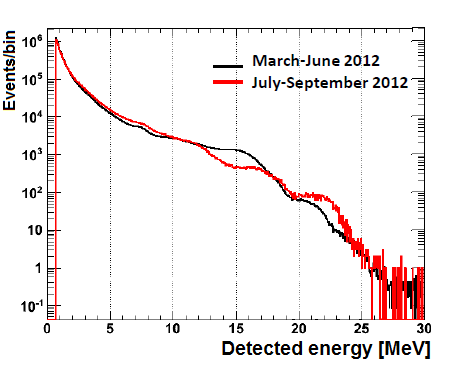}
\caption{Rate of maximum charge detection vs PMT ID number  ({\it left}), showing the units with higher rate that can be identified as possible hot \LN emitters. Detected energy spectrum of \LN  ({\it right}) for two sets of data corresponding to the first  Double Chooz publication (\textit{black}) and  to a later sample (\textit{red}). }
\label{fig:pmtqmax_en}
\end{center}
\end{figure}

The entire energy region of interest ([0.4,20] MeV) seems to be contaminated by the  \LN back\-gro\-und. At higher energies  the emitter can be more easily identified since the majority of the photons produced are detected by the PMT itself. The rate of events at which a PMT detects the maximum charge is shown in Fig.  \ref{fig:pmtqmax_en}-{\it{left}} as a function of the PMT id-number, allowing the units in the array showing the higher rate to be identified as the hottest flashers. At energies closer to the detector  threshold,  the identification is more difficult  and more complex parameters based on the light pattern and the hit time distribution have to be considered in order to identify the spurious  events  (see Sec. \ref{SecCuts}). The detected  energy spectrum after \LN selection, shown in Fig. \ref{fig:pmtqmax_en}-{\it{right}}, is falling roughly exponentially.

It has not been possible to carry out a robust study of the \LN emission position through  the vertex reconstruction algorithms since the unusual light pattern detected translates into a mis-reconstruction   of the  emission point by the  software, such that the light appears to be emitted very often from hot spots in the center of the ID.

\begin{figure}[t]
\begin{center}
\includegraphics[width=\linewidth]{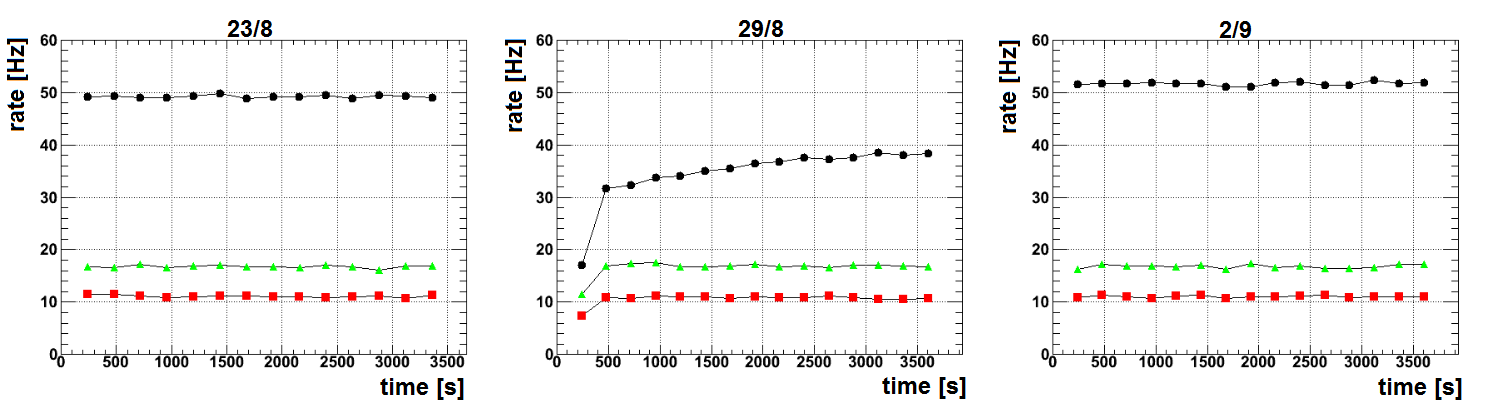}
\caption{Rate variations for three consecutive runs. The black dots correspond to \LN events, the red squares to $\bar{\nu}_e$ preselected candidates (see ref. \cite{DCG3} for the details on the selection) and the green triangles to muon events. The PMTs were switched-off after the fist run (August, $24^{th}$) and switched-on during the second one (nominal voltages reached at t $\approx$500 s - August, $29^{th}$) .}
\label{fig:switchoffhv}
\end{center}
\end{figure}

During the commissioning of the far detector, several tests were performed in order to decrease the \LN rate. The PMT voltages were lowered  such that the  gains were finally 5/6 of the nominal value. As a result the rate of {\em prompt} ($\in$[0.4,20] MeV) events decreased by $\approx$ 15 \% before applying any  rejection cuts, while the {\em delayed} ($\in$[4,10] MeV) Gd capture signal rate went down by  $\approx$ 40 \%.
Switching off PMTs for a while had only a temporary impact on the background. In Fig. \ref{fig:switchoffhv} the  rate of $\bar{\nu}_e$ candidates \cite{DCG3} is compared with the ones of  the events that are vetoed by the $\mu$ veto or  tagged as \LN (see Sec. \ref{SecCuts}) for  dedicated runs taken before and after a six days power cut.     
The IBD candidates and muons rates, after a transient given by the power up of the PMTs (t <500 s, Fig. \ref{fig:switchoffhv}-{\it{middle}}), seem to be stable and similar for all the runs, however the \LN rate decreased temporarily right after the PMTs were powered back on and went back to a value  similar  to the original one after some days.  

At the end of the commissioning phase in March 2011 the 15 hottest PMTs, identified among the more active units in events in which one PMT detected at least 10 \% of the total charge ( $q_{\textrm{max}}/q_{\textrm{tot}}> $0.1), were switched off. This resulted in the overall emission rate in the detector being reduced by $\approx$~30 Hz. Dedicated  studies, also with Monte Carlo simulations,  proved that the impact of the previous actions on the detector performance was minimal  and no substantial anisotropy  in the detector response was introduced since the switched-off PMTs were randomly distributed.
The detector configuration was adopted as the default configuration and was kept stable over the physics runs.

\subsection{Evolution of the  Light Noise during the physics runs}

\begin{figure}
\begin{center}
\includegraphics[width=0.49\linewidth]{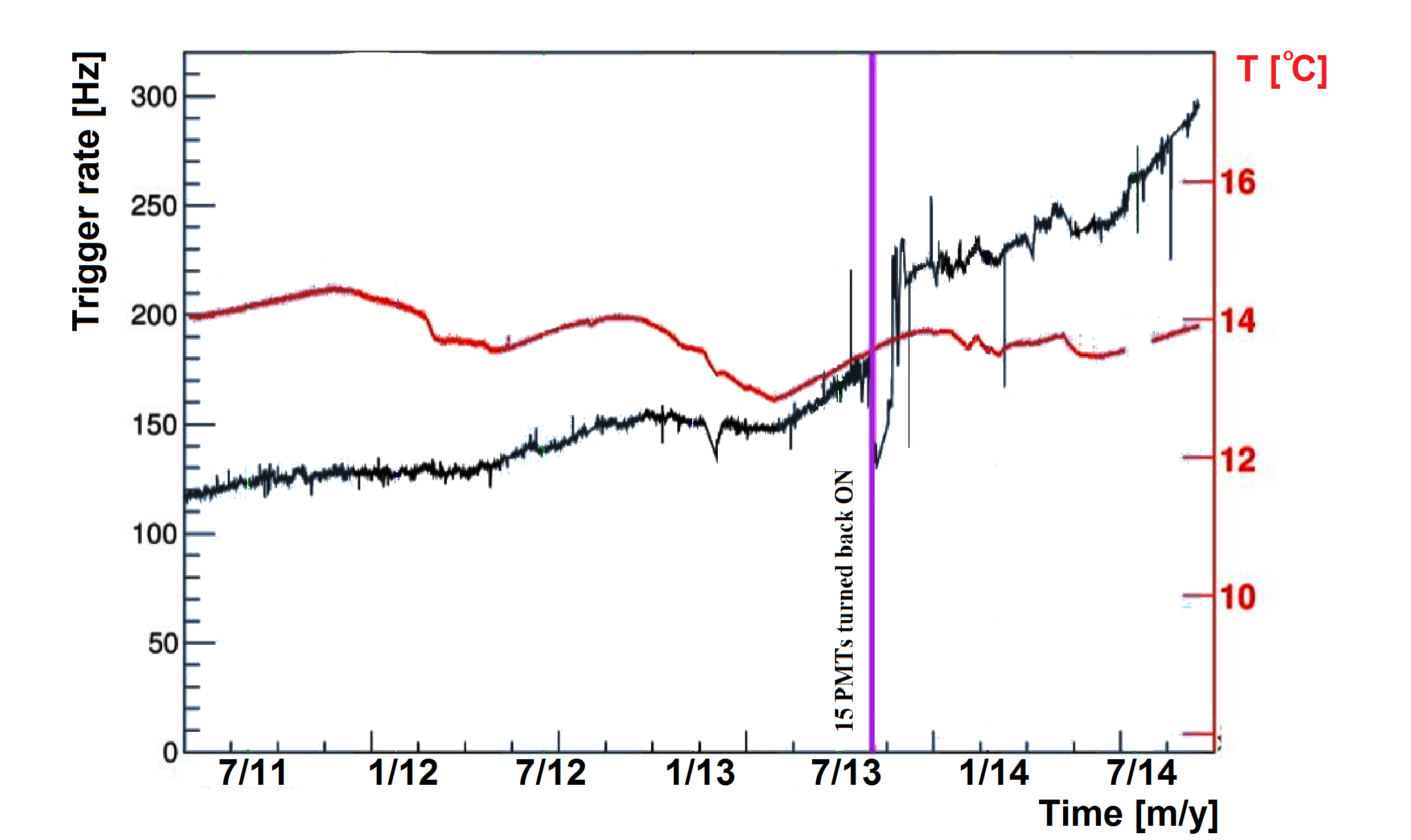}
\includegraphics[width=0.5\linewidth]{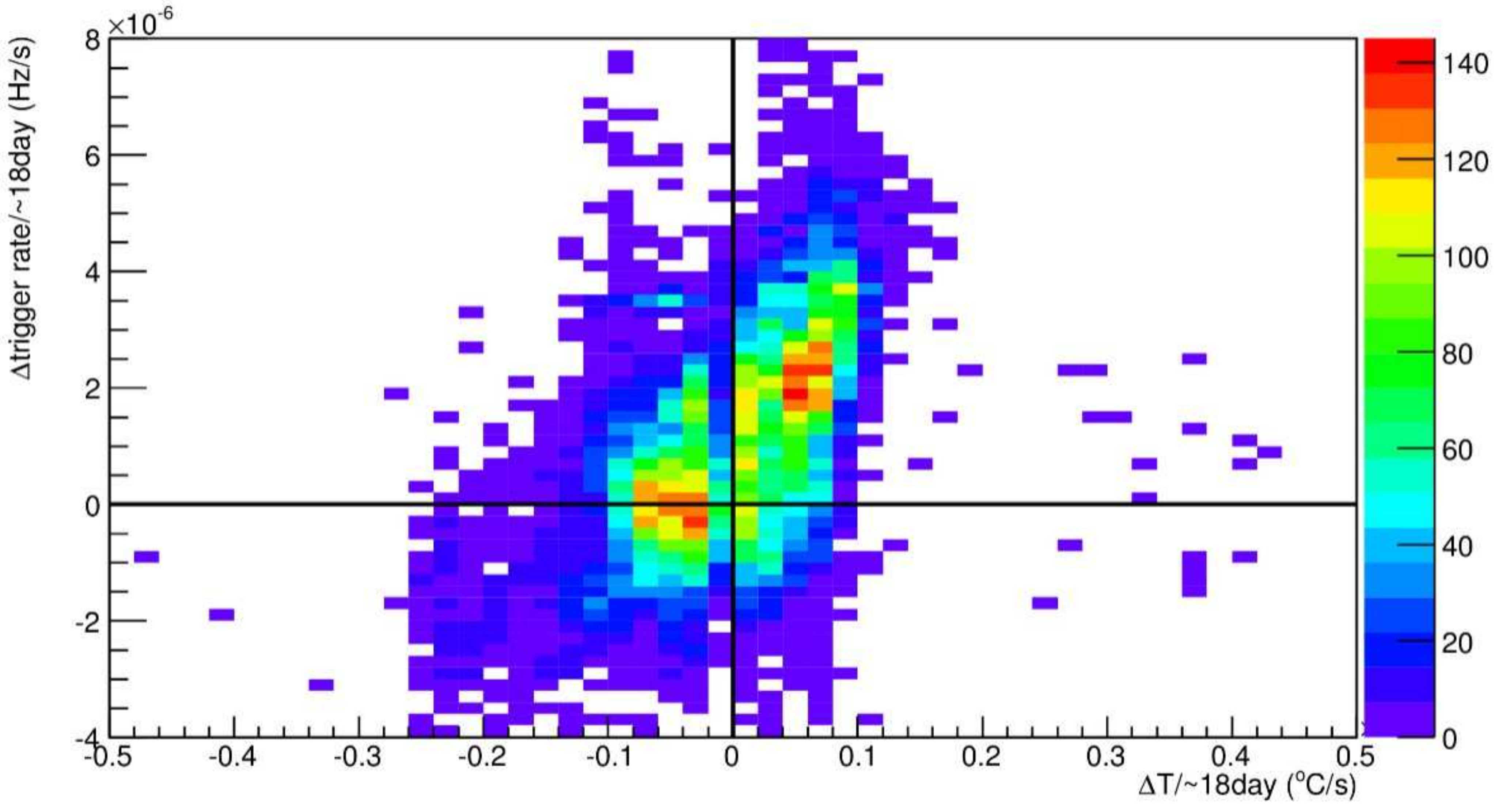}
\caption{Variation in time of the trigger rate ({\it black line}) and  buffer temperature ({\it red line}) from the start of the data taking  ({\it left}): the sudden jump in the rate was produced by the 15 hot PMTs turned back on. Correlation between trigger rate and temperature gradient ({\it right})}. 
\label{fig:correlationtemp}
\end{center}
\end{figure}

The \LN emission rate of each single  photomultiplier has been  unstable over  the data taking. Some units showed an increasing emission which, for a subset of them, returned to normal. 
In spite of the actions taken during the commissioning phase to reduce the  light  emission, the overall  trigger rate increased during the physics runs although  the  rate of the physics events was stable in the same period of time. In Fig. \ref{fig:correlationtemp}-\textit{left} the variation of the total trigger rate is shown  along with the  temperature  of the mineral oil in the  buffer volume,   evidencing a possible seasonal modulation of the background that increases from April to October while slightly decreases or remains stable from November to March.  Even if  quite small  variations of the temperature ($\approx$ 0.5 $^{\circ}$C) have been measured, a positive gradient of the temperature has a clear impact on the event rate  (Fig. \ref{fig:correlationtemp}-\textit{right}), while a corresponding decrease has not been measured  during the winter  negative gradients. As a result a total trigger rate measured at the end of  2014, after three years of operations, was more than double the initial value.  That could be explained because the rise of the temperature allows the polarization of epoxy molecules but negative temperature gradients freezes the status and do not reverse the effect, as was already observed during  the laboratory test. 

\section{Cuts and background rejection} 
\label{SecCuts}

As discussed in Sec. \ref{Sec:Comm}, it was not possible to reject the \LN events through the reconstructed vertex position, thus a set of variables have been studied in order to discriminate that  background from the neutrinos interactions. \LN pulses are characterized by an irregular shape with slower rise time with respect to the liquid scintillator pulses. A simple cut  based on the rise time of the detected signal\textsuperscript{\footnotemark[5]}\footnotetext[5]{~Sum of signals produced by all the PMTs} $t_{\textrm{rise}}$ proved to have a certain \LN identification capability. However,  in order to guarantee a significant discrimination efficiency, the pulse had to be corrected by the time of flight between the reconstructed event vertex and each PMT position. Since the $t_{\textrm{rise}}$ cut depends on the position reconstruction, we preferred to  investigate a different  low-level set of cuts that were not depending on the software algorithms.

\begin{figure}[t!]
\begin{center}
\includegraphics[width=0.48\linewidth]{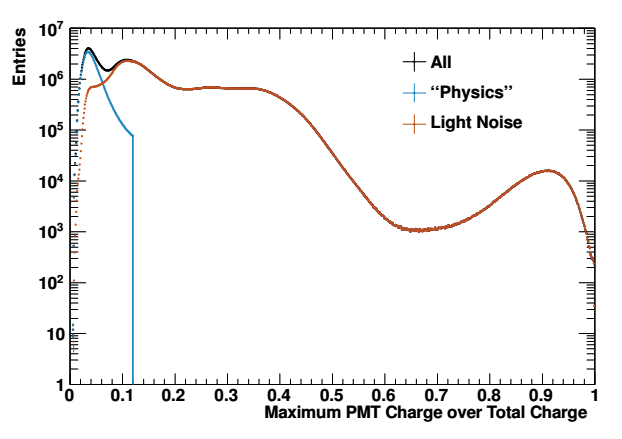}
\includegraphics[width=0.51\linewidth]{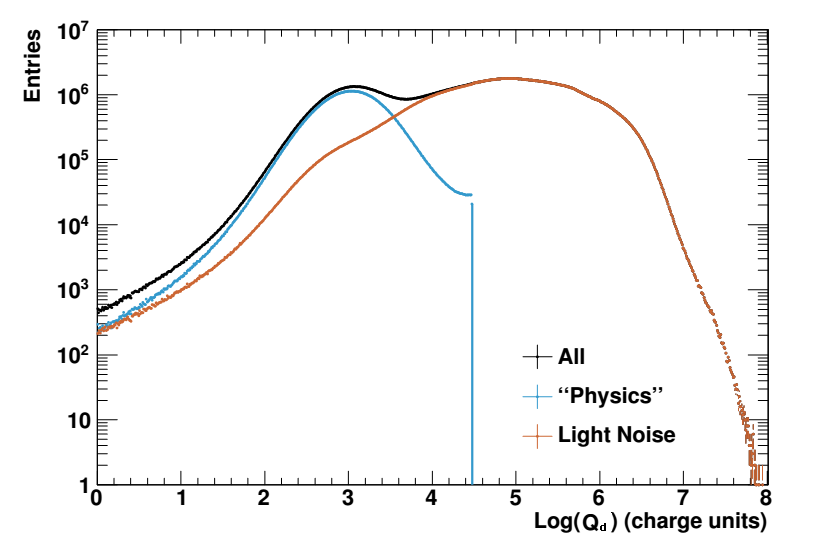}
\caption{Histograms of  $q_{\textrm{max}}/q_{\textrm{tot}}$  ({\it left}) and  of $Q_{\textrm{dev}}$  ({\it right}) for events before and after the \LN selection (see the text for details). The  impact of the cuts $q_{\textrm{max}}/q_{\textrm{tot}}<$  0.12 and  $Q_{\textrm{dev}}<$30000 sample is evidenced.} 
\label{fig:qcut}
\end{center}
\end{figure}

Events in the energy range of interest ( $E_{\textrm{vis}}\in [0.4,20]$ MeV), and with no significant activity in IV ($Q_{\textrm{IV}}<30000$ charge units, see  ref. \cite{DCG3} for the details) are considered for the analysis. The \LN emission, being localized in the PMT base, leads to inhomogeneous distributions in charge and time among all the PMT. Therefore, one time-based as well as three charge-based variables were designed  to identify and reject both the fast high energy \LN events as well as the low energy long pulses:

\begin{itemize}
  \item $q_{\textrm{max}}/q_{\textrm{tot}}$ corresponds to the ratio of the maximum charge recorded by a PMT over the total charge in the event
  \vspace{-0.25cm}
  \item $Q_{\textrm{dev}}$ is defined as $Q_{\textrm{dev}} = 1/N \times \sum_i^N \textrm{(}q_{\textrm{max}}-q_i \textrm{)}^2/q_i$ where $N$ is the number of PMTs within a sphere of 1~m radius centered at the PMT with the maximum charge
  \vspace{-0.25cm}
   \item $\sigma_{\textrm{t}}$ corresponds to the standard deviation of the PMTs hit time distribution
  \vspace{-0.25cm}
  \item $\sigma_{\textrm{q}}$ corresponds to the standard deviation of the PMTs integrated charge distribution
\end{itemize}

\begin{figure}[t!]
  \centering \includegraphics[width=.7\linewidth]{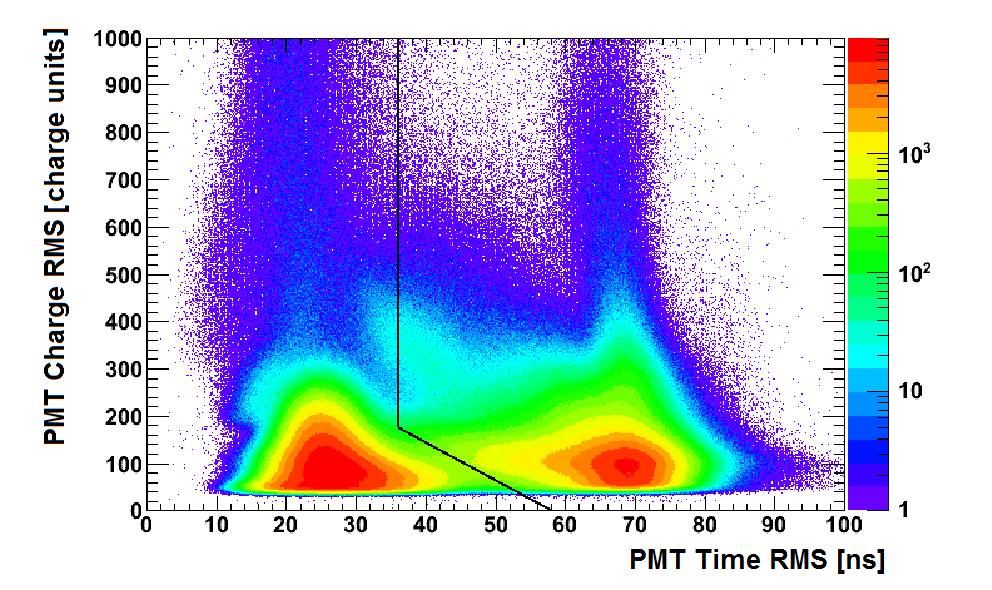}
  \caption{$\sigma_{\textrm{q}}$ vs $\sigma_{\textrm{t}}$ plane. $q_{\textrm{max}}/q_{\textrm{tot}}$ and $Q_{\textrm{dev}}$ cuts are applied. The black line represents the 2D cut on the $\sigma_{\textrm{q}}/\sigma_{\textrm{t}}$ plan. Events of interest belong to the left side of this line whereas \LN events belong to the right side of this line.}
  \label{RMSTQ_cuts}
\end{figure}

Typically the PMT  responsible for the light emission detects itself the maximum charge, thus the value $q_{\textrm{max}}/q_{\textrm{tot}}$ for those events tends to be larger than that of an interaction in the liquid scintillator. The $q_{\textrm{max}}/q_{\textrm{tot}}$ cut is then set to 0.12 (Fig. \ref{fig:qcut}-{\it left}),  the events with smaller values being considered as a possible scintillation emission inside the inner detector, while the others are classified as \LN.     $Q_{\textrm{dev}}$ represents the charge non-uniformity for PMTs around the one with maximum charge, which tends to be larger for high energy \LN events.  The $Q_{\textrm{dev}}$ variable is set to be lower than 30000 charge units for  the events of interest  (Fig. \ref{fig:qcut}-{\it right}).

Additionally, a $\bar{\nu}_e$-like event taking place inside the target is expected to have a nearly isotropic distribution of the light inside the inner detector, with a small spread in the photons arrival time at the PMTs.  \textit{A contrario}, in case of \LN  events characterized by long train of pulses emitted on a PMT base, the flasher, as well as the surrounding PMTs, detect promptly more photons  than the units in the other side of the detector. Larger values of   $\sigma_{\textrm{t}}$ are  expected for the \LN events with respect to the physics events. 

\begin{figure}[t!]
  \centering \includegraphics[width=.495\linewidth]{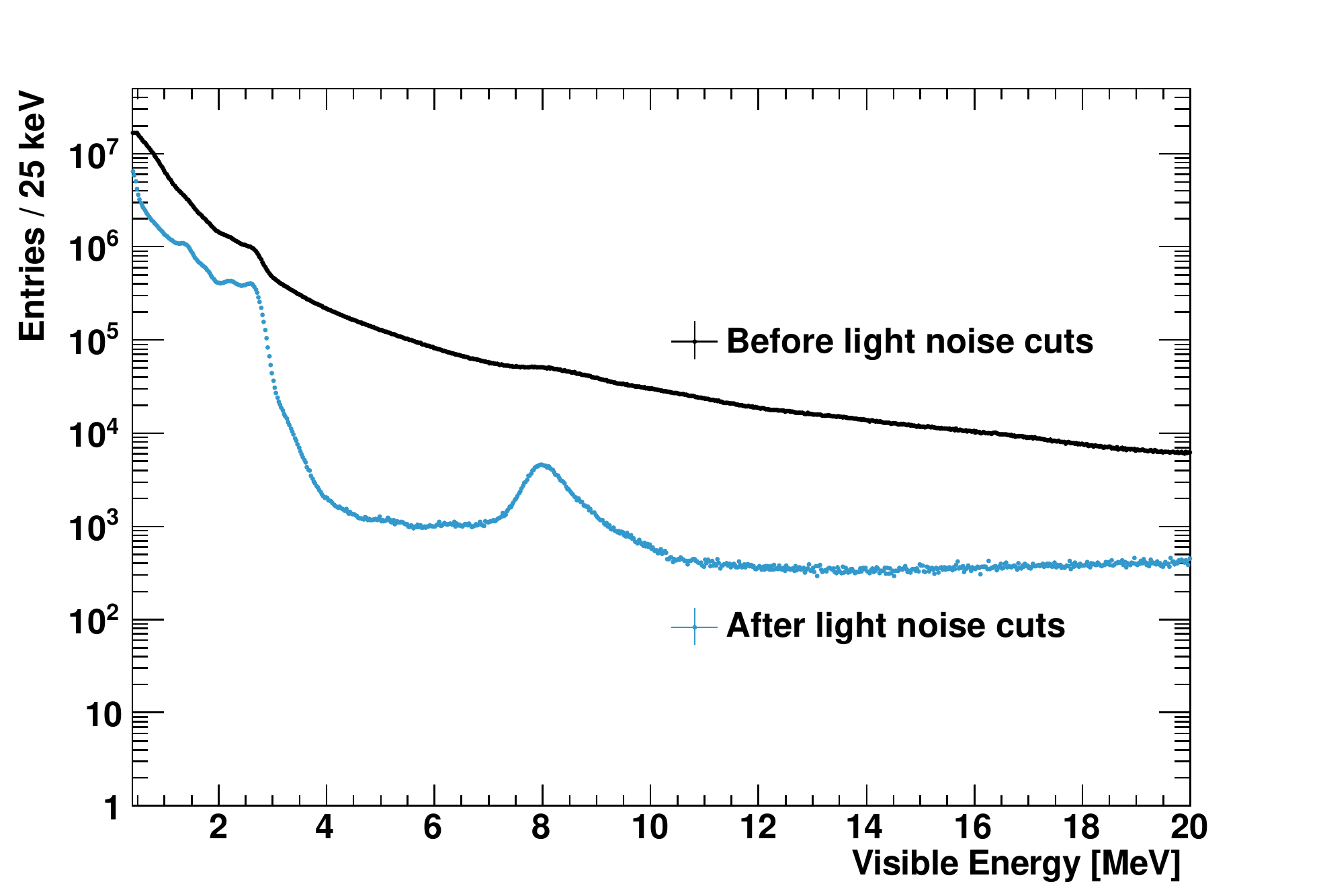}
  \centering \includegraphics[width=.495\linewidth]{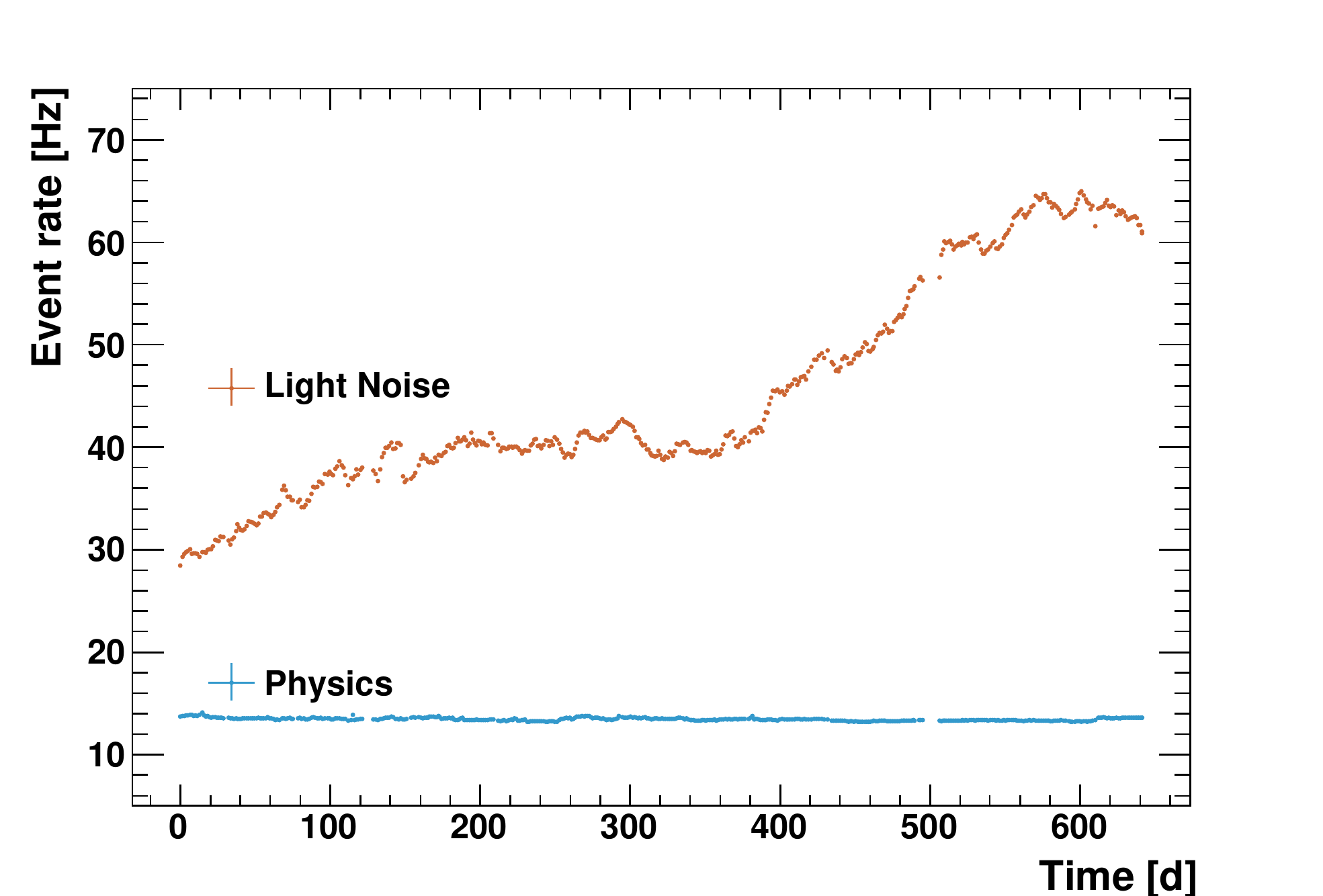}
  \caption{Visible energy spectra ({\it left}) obtained before ({\it black}) and after  ({\it blue}) the  \LN rejection cuts. Evolution of the rates over time ({\it right}) for physics  ($blue$) and \LN event  ($orange$).}
  \label{TotalCut}
\end{figure}

While the $\sigma_{\textrm{t}}$ cut rejects efficiently the \LN at higher energies,  its combination with a cut based on the standard deviation of the PMTs integrated charge distribution $\sigma_{\textrm{q}}$ proved to be more efficient in rejecting the \LN at energies $\approx$ 1 MeV.
Fig. \ref{RMSTQ_cuts} shows the $\sigma_{\textrm{q}}$ vs $\sigma_{\textrm{t}}$ spectra after applying both the $q_{\textrm{max}}/q_{\textrm{tot}}$ and $Q_{\textrm{dev}}$ cuts. Two populations are distinguishable, separated by the black line, with the events of interest in the region at low  $\sigma_{\textrm{q}}$ and $\sigma_{\textrm{t}}$ selected through $\sigma_{\textrm{t}} < 36$~ns or $\sigma_{\textrm{q}} < 464 - 8 \times \sigma_{\textrm{t}}$.  Whereas previous analyses \cite{DCG1,DCG2,DCH1} used to remove events above 40~ns, the 2D cut minimizes the detection inefficiency at  low energy  allowing to lower the analysis threshold from 0.7 MeV to 0.4 MeV \cite{DCG3}. The final cuts used to reject the \LN events are then:
\begin{itemize}
  \item $q_{\textrm{max}}/q_{\textrm{tot}} < 0.12$
  \vspace{-0.25cm}
  \item $Q_{\textrm{dev}} < 30000$ charge units
  \vspace{-0.25cm}
   \item $\sigma_{\textrm{t}} < 36$ ns or $\sigma_{\textrm{q}} < 464 - 8 \times \sigma_{\textrm{t}}$
\end{itemize}

The combination of  those  cuts    proved to be efficient  at discriminating event by event  between \LN and real interactions inside the target. The results presented here are based on the analysis of one fifth of the data used in \cite{DCG3} which extends from April 13, 2011 to January 30, 2013. As stated before, the events with visible energy not in the region of interest  for the $\bar{\nu}_e$ analysis (\textit{i.e.} between 0.4 and 20~MeV) are rejected, as well as the events with a charge deposition in the inner veto greater than 30000 charge units. Table \ref{tab:frac} summarizes the single and the combined rejection of the data sample after the \LN cuts. About 77.1 \% of the events acquired don't pass the selection and are identified as background produced by  spurious triggers.  

Fig. \ref{TotalCut}-{\it left} shows the energy spectrum of the data sample before and after the \LN cuts. The Gd capture peak around 8~MeV becomes clearly visible  after the \LN rejection, as well as the contribution given by the radioactive contamination below the 2.6~MeV peak from the $^{208}$Tl. The  remaining $LN$, eventually not rejected by the previous cuts, can be considered negligible for the $\theta_{13}$ analysis since the background is efficiently rejected requesting the coincidence between the prompt signal and delayed neutron capture. The possible  contribution  given by the coincidence of two uncorrelated \LN signals  is included in the accidental background, which is precisely measured through a pure accidental  sample obtained by a delayed off-time ($\textgreater$ 1 s)  window  \cite{DCG3}.  Fig. \ref{TotalCut}-{\it right} shows the event rate for the two populations selected from Fig. \ref{RMSTQ_cuts}.  As expected the physics remains stable whereas the \LN is increasing since the beginning of the data taking.  No  evidence of a correlated background produced by multiple triggers  from long \LN pulses has been found. Finally,  the preliminary comparison between the physics sample after cuts with the data taken with  near detector, whose PMT bases have been blackened (see Sec. \ref{ND}), confirms the success of the \LN rejection. 

\begin{figure}[t!]
  \centering \includegraphics[width=.49\linewidth]{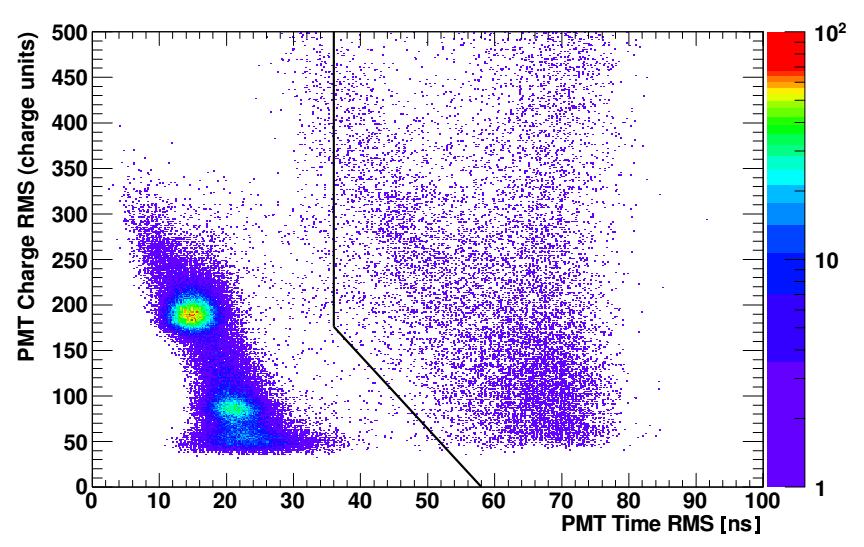}	
  \centering \includegraphics[width=.48\linewidth]{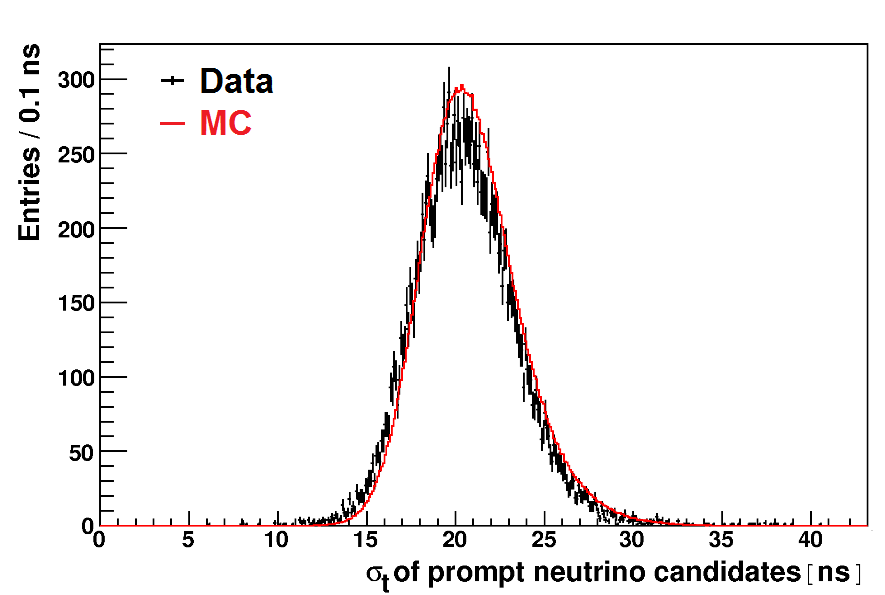}
  \caption{$^{252}$Cf  calibration data  (source at the center of the detector) without any \LN selection (\textit{left}): the Gd and the H capture peaks ($\sim$190 and at $\sim$80  charge units) can be discriminated from  the \LN events (population on the right) by the 2D cut in the  $\sigma_{\textrm{q}}$ vs $\sigma_{\textrm{t}}$  plane.  Distribution of the \LN variable $\sigma_{\textrm{t}}$ for neutrino candidates found for data and Monte Carlo  (\textit{right}).}
  \label{LNEff}
\end{figure}

{
\renewcommand{\arraystretch}{1.125}
\begin{table}[t!]
  \vspace{0.125cm}
  \begin{tabular}{ccc}
    \hline
    \hline
    \textbf{Cut} & \textbf{Single cut rejection} & \textbf{Combined rejection}\\
    \hline
    $\sigma_{\textrm{q}}/\sigma_{\textrm{t}}$ & 67.8~\% & 67.8~\%\\
    $q_{\textrm{max}}/q_{\textrm{tot}}$ & 51.9~\% & 76.2~\%\\
    $Q_{\textrm{dev}}$ & 51.4~\% & 77.1~\%\\
    \hline
    \hline
  \end{tabular}
   \centering
  \caption{Fraction of events rejected by applying \LN cuts. The first column represents the total events rejected by  each cut independently, the second column  the total events rejected by successively applying the  cuts: 22.9 \% of the events pass the selection.  }
    \label{tab:frac}
\end{table}
}

\begin{figure}[t!]
  \begin{minipage}[t!]{.49\linewidth}
    \centering \includegraphics[width=\linewidth]{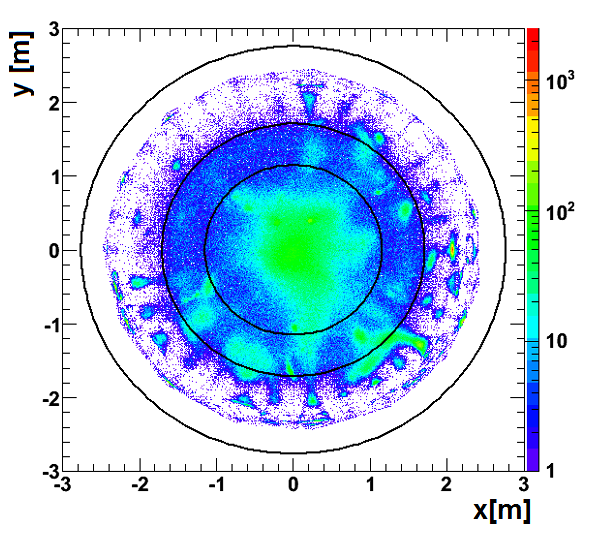}
  \end{minipage}
  \begin{minipage}[h!]{.02\linewidth}
    \centering
  \end{minipage}
  \begin{minipage}[t!]{.49\linewidth}
    \centering \includegraphics[width=\linewidth]{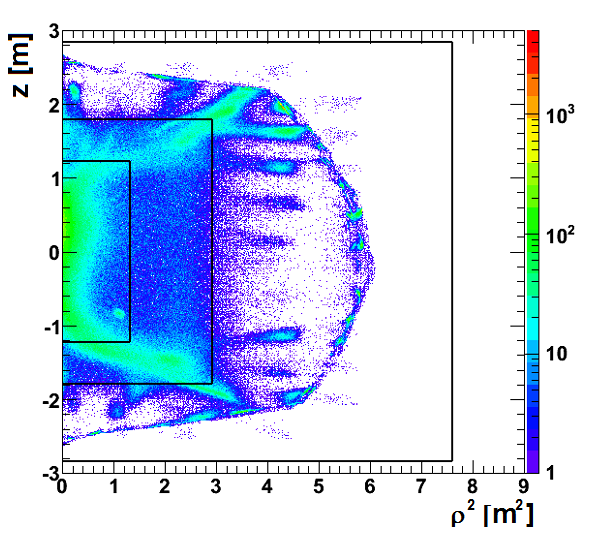}
  \end{minipage}
  \caption{ Distribution of the events on the $xy$ plane  (\textit{left}) and on the $\rho^2z$ plane (\textit{right}) from the special runs dedicated to the \LN study. \LN cuts are not applied. The black lines correspond to the target, $\gamma$-catcher and buffer.}
  \label{Ch4:rhoz}
  \vspace{1cm}
  \begin{minipage}[t!]{.49\linewidth}
    \centering \includegraphics[width=\linewidth]{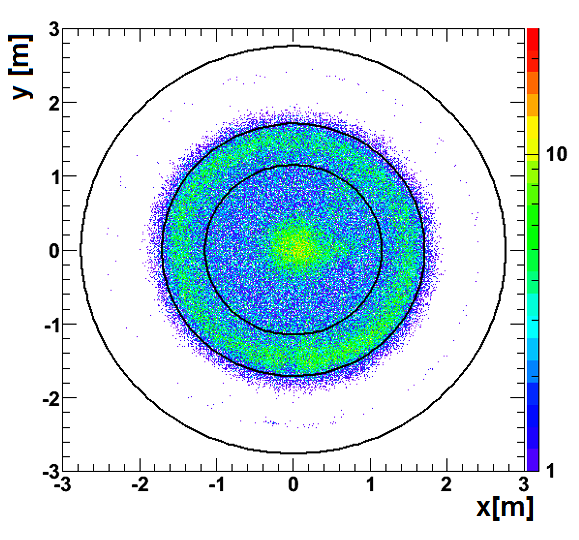}
  \end{minipage}
  \begin{minipage}[t!]{.02\linewidth}
    \centering
  \end{minipage}
  \begin{minipage}[t!]{.49\linewidth}
    \centering \includegraphics[width=\linewidth]{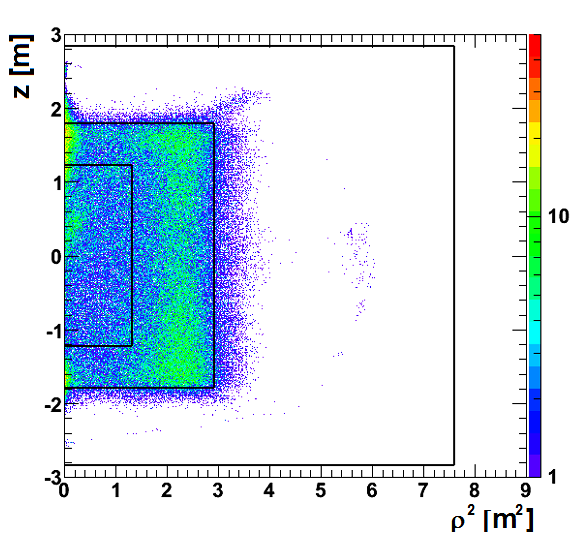}
  \end{minipage}
  \caption{Distribution of the events on the $xy$ plane (\textit{left}) and on the $\rho^2z$ plane (\textit{right}) from the special runs dedicated to the \LN study, after applying the cuts.}
  \label{Ch4:rhoz_cuts}
\end{figure}

The \LN strategy has been validated firstly on the calibration data. Different neutron calibration runs with a $^{252}$Cf source placed at different positions along $z$-axis have been analyzed. In Fig. \ref{LNEff}-{\it{left}} the events  reconstructed in a sphere of 50 cm radius  around the source are plotted in the   $\sigma_{\textrm{q}}$ vs $\sigma_{\textrm{t}}$  plane after the same  $E_{\textrm{vis}}$  and $Q_{\textrm{IV}}$ cuts detailed before. We clearly see a good separation between the physics events on the left and the \LN on the right, with the  the Gd capture peak, visible at ~190 charge units, not affected by the 2D cut. Dedicated simulations have been carried out in order to tune the rejection  criteria (Fig. \ref{LNEff}-{\it{right}}) and estimating the inefficiency on the inverse beta decay selection.  The Monte Carlo based calculation of the inefficiency on IBD candidate was evaluated to be at $0.0004 \pm 0.0002~\%$ for $\sigma_{\textrm{q}}/\sigma_{\textrm{t}}$ cut, $0.0118 \pm 0.0008~\%$ for $q_{\textrm{max}}/q_{\textrm{tot}}$ and $0.0004 \pm 0.0002~\%$ for $Q_{\textrm{dev}}$. Combining these 3  rejection cuts, the total detection inefficiency is $0.0124 \pm 0.0008~\%$, that can be considered negligible for the neutrino oscillation analysis.

Since the beginning of the data taking, some noisy PMTs were turned off in order to reduce the spurious trigger contamination. A few hours of dedicated runs with these PMTs turned back on were taken later on for further \LN studies. Fig. \ref{Ch4:rhoz} shows the spatial distribution of the events for these dedicated runs before applying the \LN cuts. One can clearly see structures in the buffer, the $\gamma$-catcher and the target, however it would be very difficult to identify the noisy PMTs through the position reconstruction. Fig. \ref{Ch4:rhoz_cuts} shows the same distributions after applying the \LN cuts: only a few events remain reconstructed in the buffer and a smooth distribution, without clear artifacts, is evident in the neutrino target. The success of the \LN strategy allowed to turn these noisy PMTs back on in 2013.

\section{The near detector}   
\label{ND}

In order to reduce the trigger rate caused by the $LN$, the PMT bases of the near detector were covered with a  radiopure and chemically-compatible polyester film (TORAY, Lumirror X30 \cite{toray}) which is the same type used in the KamLand experiment.

\begin{figure}[t!]
\begin{center}
  \includegraphics[height=.3\textheight]{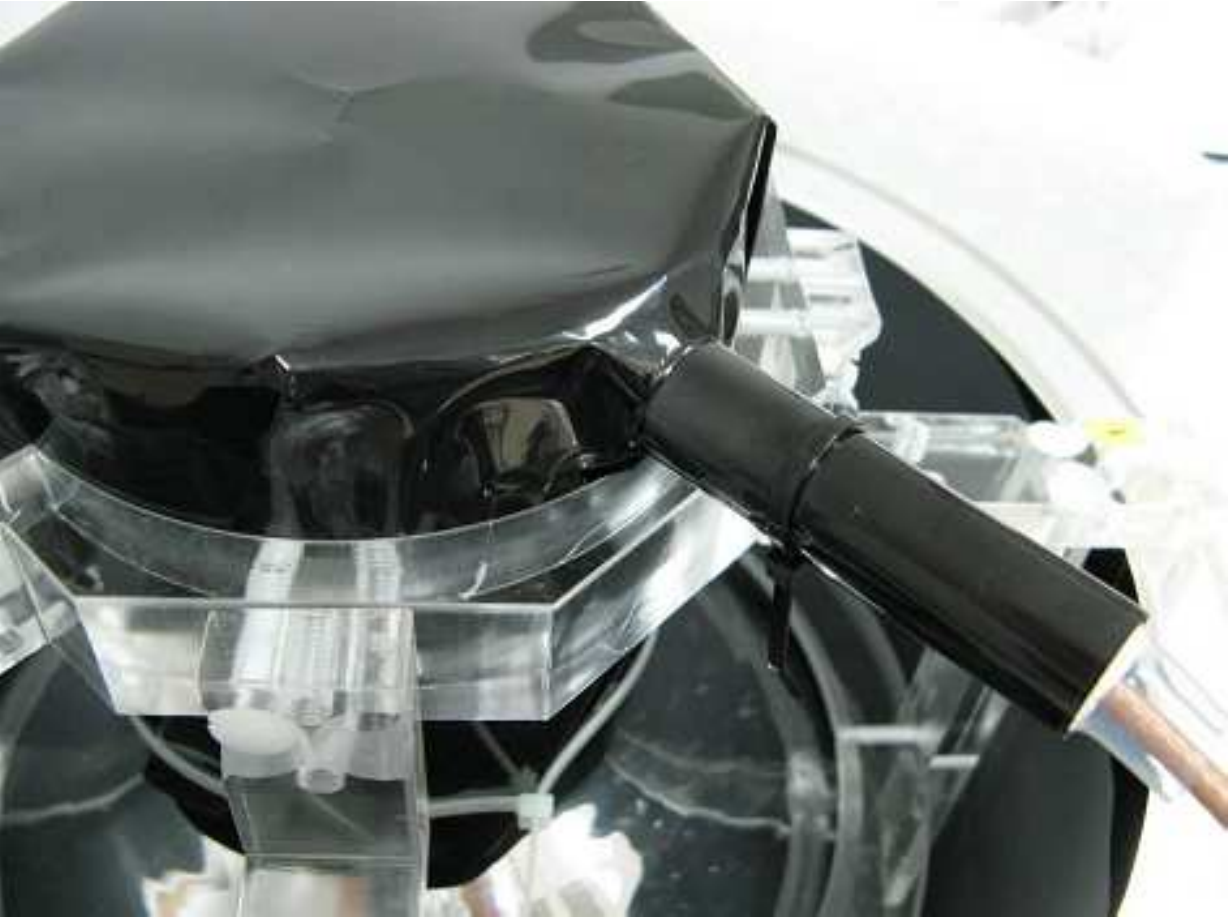}
  \caption{PMT base covered with black sheet as made for the near detector.}
  \label{fig:cover}
  \end{center}
\end{figure}

Several designs and different thicknesses of  the sheet were tested, measuring  the corresponding reduction in the \LN trigger rate using the laboratory setup described in Sec. \ref{SecTest}. The measurements were carried out at room temperature ($29^{\circ} \rm{C}$) with a flasher PMT high voltage set at 1900 V, higher than normal operating high voltage in order to maximize the light emission. The highest reduction (99.88 \%) was found for the design consisting of a single piece of sheet folded around the PMT base, with a single cut ending in a hole to let the cable exit (Fig. \ref{fig:cover}). The base cover was attached to the PMT mechanically, using the PMT support structure, avoiding the use of glue. The cable exit had a rectangular strip of sheet wrapped around it to prevent the light escaping through it. Similar results were found for the 80 $\mu \rm{m}$ and 100 $\mu \rm{m}$ thicknesses. We chose the thinner sheet since it was easier to fold.  Small holes were punctured in the covers to allow the air to escape during the detector filling process, thus avoiding the formation of trapped air bubbles inside the detector.

A spectroscopy study of the optical properties    of the   $80 \mu \rm{m}$ sheet was carried out independently, measuring its  transmittance as a function of the wavelength.  While the transmittance is practically null in the UV region, it slowly rises in the IR, being on average of the order of 0.16 \%   in the visible region, in agreement with the reduction measured in the laboratory setup. The bases of all the 390 PMTs for the Inner Detector were covered with the black sheet.

\begin{figure}[t!]
\begin{center}
  \includegraphics[height=.3\textheight]{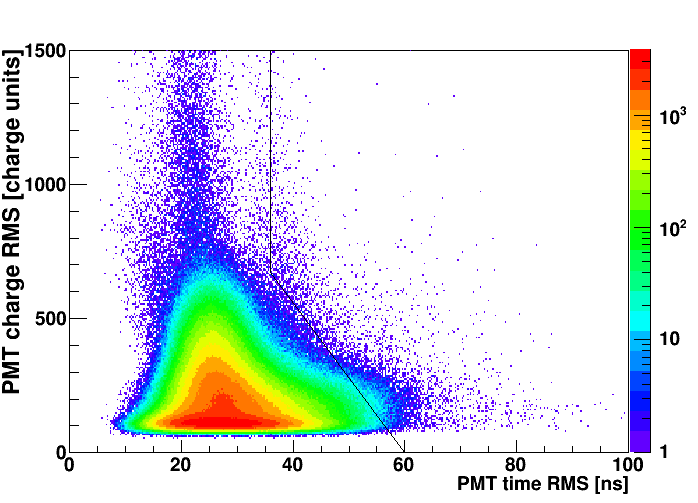}
  \caption{$\sigma_{\textrm{q}}$ vs $\sigma_{\textrm{t}}$ spectra of the first ND data. The black line represents a 2D cut, similar to the one used for the FD analysis (see Fig. \ref{RMSTQ_cuts}), adjusted taking into account the different PMTs gain between FD and ND (see text for details).}
  \label{fig:ND}
  \end{center}
\end{figure}

The covering of the base reduced drastically the \LN in the near detector.  Fig. \ref{fig:ND} shows that only one population of events is evident in the $\sigma_{\textrm{q}}$ vs $\sigma_{\textrm{t}}$ spectra. Those events can be selected after applying $q_{\textrm{max}}/q_{\textrm{tot}}$ and $Q_{\textrm{dev}}$ cuts as it has been done for the far detector, once taken into account the different PMTs gain between the two detectors. The evidence proves that the instrumental background was correctly identified and understood.  

It is possible that a small  fraction of the light, which propagates inside the PMT,  exits through its  transparent window. The pulse can be detected by the tubes in front of the flasher. However that signal should be more easily rejected by the $q_{\textrm{max}}/q_{\textrm{tot}}$ cut described in Sec.  \ref{SecCuts}, making the impact of the $LN$, on both physics data and trigger rate, negligible in the near detector. 
Indeed, no clear evidence of the \LN is present in the first near detector data,  although  the detector behavior will be carefully monitored  during the entire data taking looking for possible hints of \LN.
 
\section{Conclusions}

The emission of light inside the  optical volume has been observed during the preliminary operations of the Double Chooz far detector in the form of fast ($\approx{10-100}$ ns) flash of light  or of a train  ( $\approx{}1-10$ $\mu$s) of  signals similar to the glowing in gas. Even though only one PMT typically detects a large pulse, the light spreads out among the other PMTs after several reflections  triggering the DAQ and contaminating the physics data sample. 
Tests in-situ and in an external laboratory proved that the ensemble of phenomena called {\textit{Light Noise}} was produced by the combined effect of heat and high voltage on the epoxy resin covering  the photomultiplier bases. The $Light~Noise$ rate rose during the  data taking with the far detector, such that more than 75 \% of the total triggers  are currently produced by the spurious light emission. A correlation between the light emission rate and the temperature of the liquid  scintillator was found, with a seasonal variation of the rate gradient.

 Since the acquired data  were contaminated by this instrumental background,   a set of cuts was  used to select the physics events in the off-line analysis. One time-based as well as three charge-based variables were designed  to identify and reject the different types of $Light~Noise$ detected at low as well as high energy, with an  estimated combined signal inefficiency of 0.0124$\pm$0.0008 \%, which was considered negligible for the neutrino oscillation analysis.

At the same time the  bases of the near detector PMTs have been covered with a black sheet of a radiopure material in order to reduce the impact of this background on the total trigger rate. The modification has drastically reduced the contamination of the spurious events in the physics data, and  is expected to improve further the efficiency of the analysis cuts. The evolution of the $Light~Noise$ rate and  energy spectrum is constantly monitored during the data taking of the experiment with both detectors in operation.

Several other experiments reported similar instrumental effects. It is not possible to directly extrapolate our results to other detectors since no systematic study or explanation of the light emission mechanism  have been published. Nonetheless, the results of the  investigations can be particularly relevant  for experiments that are known to use similar epoxy resins or base assembly.

\section*{Acknowledgments}

We thank the French electricity company EDF; the
European fund FEDER; the R\'{e}gion de Champagne de Champagne Ardenne; the D\'{e}partement des Ardennes; and the Communaut\'{e} de Communes Ardenne Rives de Meuse.  We acknowledge the support of the CEA, CNRS/IN2P3, the computer centre CCIN2P3, and LabEx UnivEarthS in France (ANR-11-IDEX-0005-02); the Ministry of Education, Culture, Sports, Science and Technology of Japan (MEXT) and the Japan Society for the Promotion of
Science (JSPS); the Department of Energy and the National Science Foundation of the United States; U.S.
Department of Energy Award DE-NA0000979 through the Nuclear Science and Security Consortium; the Ministerio de Econom\'{i}a y Competitividad (MINECO) of Spain; the Max Planck Gesellschaft, and the Deutsche
Forschungsgemeinschaft DFG, the Transregional Collaborative Research Center TR27, the excellence cluster ``Origin and Structure of the Universe'', and the Maier-Leibnitz-Laboratorium Garching in Germany; the Russian Academy of Science, the Kurchatov Institute and RFBR (the Russian Foundation for Basic Research); the
Brazilian Ministry of Science, Technology and Innovation (MCTI), the Financiadora de Estudos e Projetos
(FINEP), the Conselho Nacional de Desenvolvimento Cient\'{i}fico e Tecnol\'{o}gico (CNPq), the S\~{a}o Paulo Research Foundation (FAPESP), the Minas Gerais State Research Foundation (FAPEMIG, project CEX-APQ-01439-14), and the Brazilian Network for High Energy Physics (RENAFAE) in Brazil.

\end{document}